\documentclass[useAMS,usenatbib]{mn2e}
\usepackage{epsfig}

\newcommand\figwidth{1.01}
\newcommand\figwidthCB{0.7}
\def\Mpc{{\rm\thinspace Mpc}}

\def\s{{\rm\thinspace s}}
\def\km{{\rm\thinspace km}}
\def\kms{\hbox{$\km\s^{-1}\,$}}
\def\gtrsim{\mathrel{\hbox{\rlap{\hbox{\lower4pt\hbox{$\sim$}}}\hbox{$>$}}}}
\def\lesssim{\mathrel{\hbox{\rlap{\hbox{\lower4pt\hbox{$\sim$}}}\hbox{$<$}}}}

\title[The WiggleZ Dark Energy Survey]{The WiggleZ Dark Energy Survey: Survey
  Design and First Data Release}
\author[Drinkwater et al.]{Michael J.\
Drinkwater$^{1}$\thanks{E-mail: m.drinkwater@uq.edu.qu}, 
Russell J.\ Jurek$^{1}$, Chris Blake$^{2}$, David
Woods$^{3,4}$,\newauthor
Kevin A.\ Pimbblet$^{1}$, 
Karl Glazebrook$^{2}$, Rob Sharp$^{5}$,
Michael B.\ Pracy$^{2,6}$,\newauthor
Sarah Brough$^{2}$, Matthew Colless$^{5}$,
Warrick J.\ Couch$^{2}$, Scott M.\ Croom$^{7}$,\newauthor
Tamara M.\ Davis$^{1,8}$, Duncan Forbes$^{2}$, Karl Forster$^{9}$, David
G.\ Gilbank$^{10}$,\newauthor
Michael Gladders$^{11}$, Ben Jelliffe$^{7}$,
Nick Jones$^{2}$, I-hui Li$^{2}$, Barry Madore$^{12}$,\newauthor
D.\ Christopher Martin$^{9}$,
Gregory B.\ Poole$^{2}$, Todd Small$^{9}$,  
Emily Wisnioski$^{2}$, \newauthor
Ted Wyder$^{9}$, H.K.C.\ Yee$^{13}$\\
$^{1}$Department of Physics, The University of Queensland, Brisbane, QLD 4072, Australia\\
$^{2}$Centre for Astrophysics \& Supercomputing, Swinburne University
of Technology, P.O.\ Box 218, Hawthorn, VIC 3122, Australia\\
$^{3}$Department of Physics \& Astronomy, University of British Columbia, 6224 Agricultural Road, Vancouver, B.C., V6T 1Z1, Canada\\
$^{4}$School of Physics, University of New South Wales, Sydney, NSW 2052,
Australia\\
$^{5}$Anglo-Australian Observatory, P.O. Box 296, Epping, NSW 2121, Australia\\
$^{6}$Research School of Astronomy \& Astrophysics, Australian
National University, Weston Creek, ACT 2600, Australia \\
$^{7}$Sydney Institute for Astronomy, School of Physics, University of Sydney, NSW 2006, Australia \\
$^{8}$ Dark Cosmology Centre, Niels Bohr Institute, University of Copenhagen, 
DK-2100 Copenhagen, Denmark\\
$^{9}$California Institute of Technology, MC 405-47, 1200 East California Boulevard, Pasadena, CA 91125, United States \\
$^{10}$Department of Physics and Astronomy, University Of Waterloo, Waterloo, Ontario N2L 3G1, Canada\\
$^{11}$Department of Astronomy and Astrophysics, The University of Chicago,
5640 S. Ellis Ave, Chicago, IL 60637, United States of America\\
$^{12}$Observatories of the Carnegie Institute of Washington, 813 Santa Barbara St., Pasadena, CA 91101, United States of America \\
$^{13}$Department of Astronomy \& Astrophysics, University of Toronto, 50 St. George St., Toronto, ON, M5S 3H4, Canada
}

\begin{document}

%\date{Accepted ?. Received ?; in original form 2009 July 14}
%\date{DRAFT 1.0 submitted 2009 July 14}
%\date{DRAFT 2.0 re-submitted 2009 September 19}
\date{Accepted 2009 September 21}

\pagerange{\pageref{firstpage}--\pageref{lastpage}} \pubyear{2009}

\maketitle

\label{firstpage}

\begin{abstract}
  The WiggleZ Dark Energy Survey is a survey of 240,000 emission line
  galaxies in the distant universe, measured
  with the AAOmega spectrograph on the 3.9-m Anglo-Australian
  Telescope (AAT). The primary aim of the survey is to precisely measure 
  the scale of baryon acoustic oscillations (BAO) imprinted on the spatial
  distribution of these galaxies at look-back times of 4--8\,Gyrs.

  The target galaxies are selected using ultraviolet photometry from
  the GALEX satellite, with a flux limit of $NUV<22.8$ mag. We also
  require that the targets are detected at optical wavelengths,
  specifically in the range $20.0<r<22.5$\,mag. We use the Lyman break
  method applied to the ultraviolet colours, with additional optical
  colour limits, to select high-redshift galaxies. The galaxies
  generally have strong emission lines, permitting reliable redshift
  measurements in relatively short exposure times on the AAT. The
  median redshift of the galaxies is $z_{med}=0.6$. The redshift range
  containing 90\% of the galaxies is $0.2 < z < 1.0$.

  The survey will sample a volume of $\sim$1\,Gpc$^3$ over a projected
  area on the sky of 1,000 degree$^2$, with an average target density
  of 350\,degree$^{-2}$.  Detailed forecasts indicate the survey will
  measure the BAO scale to better than 2\% and the tangential and
  radial acoustic wave scales to approximately 3\% and
  5\%, respectively. Combining the WiggleZ constraints with
  existing cosmic microwave background measurements and the latest
  supernova data, the marginalized uncertainties in the cosmological
  model are expected to be $\sigma(\Omega_m)=0.02$ and
  $\sigma(w)=0.07$ (for a constant $w$ model). The WiggleZ measurement
  of $w$ will constitute a robust, precise and independent test of
  dark energy models.

  This paper provides a detailed description of the survey and its
  design, as well as the spectroscopic observations, data reduction,
  and redshift measurement techniques employed. It also presents an
  analysis of the properties of the target galaxies, including
  emission line diagnostics which show that they are mostly extreme
  starburst galaxies, and {\em Hubble Space Telescope} images, which
  show they contain a high fraction of interacting or distorted
  systems. In conjunction with this paper, we make a public data
  release of data for the first 100,000 galaxies measured for the
  project.

\end{abstract}

\begin{keywords}
surveys -- galaxies: high-redshift -- 
galaxies: photometry -- galaxies: starburst -- 
cosmology: observations --
ultraviolet: galaxies
\end{keywords}

%\newpage

\section{Introduction}

% {\em Which email corrections I have done: all to Karl 26/5/09, 10am.
% Remaining items:
% 
% Karl: 'random' spectra plots need minor changes and QSOs adding.
% 
% David (passed on to Warrick for polishing in case he agrees)
% Sec. 1, para. 3, sents. 9 and 10 (last two sentences):
% change "supernova" to be "supernovae" throughout these two sentences... 
% it should be the plural adjective...
% }

A major theme of cosmology for several decades has been the
determination of the best cosmological model to describe our
Universe. By the early 1990s, evidence was already accumulating from
measurements of large-scale structure and the cosmic microwave
background that if the Universe was flat, the matter density was well
below the critical value, requiring an additional (large) contribution
from a non-zero cosmological constant term
\citep{Efstathiou1990}. This ``$\Lambda$CDM''cosmology, dominated by
the cosmological constant ($\Lambda$) and cold dark matter (CDM),
although radical at the time, predicted an older age for the Universe
than other models, so was more consistent with observational lower
limits to the age \citep{Ostriker1995}. Additional early support for a
non-zero cosmological constant was provided by
\citet{Yoshii1991,Yoshii1995} in their analysis of faint near-infrared
galaxy counts. The case for a $\Lambda$CDM cosmology was strengthened dramatically in
the late 1990s by the measurement from distant type Ia supernovae of
the acceleration in the expansion rate of the Universe
\citep{Riess1998,Perlmutter1999}. 

The goal of cosmology now is to test
different models of the underlying physics responsible for this
acceleration, dubbed ``dark energy''. This research has focused around
the dark energy equation of state (parameterised by $w(z)$, so
pressure equals $w$ times density) and how it changes with time. A
central goal is to measure $w(z)$ at multiple redshifts to test
various dark energy models \citep{Frieman2008}. 
A powerful way to test the dark energy models is to measure the
geometrical relations between distance and redshift. These can be
measured by observing the baryon acoustic oscillation (BAO) scale
imprinted on the distribution of baryonic matter at recombination
\citep{Cooray2001,Eisenstein2002,Blake2003,Seo2003,Hu2003,Linder2003,Glazebrook2005}.
The BAO signal has been clearly measured in the cosmic microwave
background \citep[i.e.\ at recombination;][]{Bennett2003}. More
recently, the 2dF Galaxy Redshift Survey;
\citep[2dFGRS,][]{Colless2001} and the Sloan Digital Sky Survey
\citep[SDSS,][]{York2000} have permitted the detection of the BAO
signal in the distribution of low-redshift ($z\approx 0.2$--0.35)
galaxies. The 2dFGRS detection was reported by \citet{Cole2005} and
\citet{Percival2007}; the SDSS result was reported by
\citet{Eisenstein2005}, \citet{Huesti2006} and
\citet{Percival2007}. These galaxy samples are too nearby to be
sensitive to the cosmic effects of dark energy, but provide a superb
validation of the concept. 

Measurement of the BAO scale in the galaxy
distribution at high redshifts provides a powerful test of the dark
energy models. It gives a geometrical measurement of the Universe at
different redshifts which delineates the cosmic expansion history with
high accuracy and traces the effects of dark energy
\citep{Blake2003}. It also provides a unique and direct measurement of the
expansion rate at high redshift \citep{Glazebrook2005}. The most
important advantage of the BAO measurement is that its measurement is
completely independent of the supernova results and, in particular, is
less susceptible to systematic errors. It is also complementary to the
supernova results in that the measurement uncertainties are orthogonal
to those of the supernova measurements \citep{Blake2009a}.

% In 2006 February the Anglo-Australian Observatory (AAO) issued a call
% for very large observing proposals. In preparation for that, workshops
% were held in both the UK (at University College, London, in 2004 June)
% and in Australia (at the the AAO Headquarters, Epping, in 2005
% February) to discuss possible extra-galactic projects for the new
% AAOmega multi-object spectrograph \citep{Smith2004,Saunders2004}. The
% WiggleZ survey \citep{Glazebrook2007} grew out of a proposal presented
% at both workshops for a large BAO experiment at higher redshifts than
% the SDSS and 2dFGRS detections. At the time no other observatories
% world-wide had suitable instrumentation to carry out such a
% project. The WiggleZ project was subsequently approved in 2006 May as
% the first of the large projects for the AAO, and observations started
% in 2006 August.
 
The WiggleZ Dark Energy Survey is a major large-scale structure survey
of UV-selected emission-line galaxies.  The survey is designed to map
a cosmic volume large enough to measure the imprint of baryon
oscillations in the clustering pattern \citep[this requires $\sim 1$
Gpc$^3$;][]{Blake2003} at a significantly higher redshift ($0.2<z<1.0$)
than has been previously achieved.  Its core
science goals are: \vspace*{-2mm}
\begin{enumerate}
\parskip 0pt
\parsep 0pt
\parindent 1cm
\item To calculate the power spectrum of the galaxy distribution with
  sufficient precision to identify and measure the baryon acoustic
  oscillation scale to within 2 per cent;
\item To determine the value of the dark energy equation of state
  parameter, $w$, to within 10 per cent when combined with CMB data, and
  hence either independently validate or refute the currently-favoured
  ``cosmological constant'' model.
\end{enumerate}\vspace*{-2mm}
The redshift range of the WiggleZ Dark Energy Survey will make it the
first BAO survey to sample the epoch when the universe makes the
transition from matter-dominated to lambda-dominated expansion.

In this paper we present a detailed description of the survey and 
the key innovations that were developed to allow it to proceed. 
We start with a description of the survey design and its requirements in
Section~\ref{sec-design}.  This is followed in
Section~\ref{sec-target} with details of the selection of our
target galaxies from UV data combined with optical data.  The UV
selection of the target galaxies means they have strong emission
lines, so their redshifts can be measured in relatively short
exposures on a 4m-class telescope like the Anglo-Australian Telescope
(AAT).  In Section~\ref{sec-spectroscopy} we describe our AAT
spectroscopic observations and data reduction. The processing of our
data to obtain reliable spectroscopic redshifts is detailed in
Section~\ref{sec-redshifts}. This includes analysis of the reliability
of our redshifts. In Section~\ref{sec-results}, we present a summary
of the initial results of the survey, notably the galaxy properties
and the success of the survey design.  Following this, in
Section~\ref{sec-data} we describe the data products of the survey, as
released in our first public data release accompanying this paper. We
summarise this paper and outline the remaining stages of the survey in
Section~\ref{sec-summary}.

A standard cosmology of $\Omega_m = 0.3$, $\Omega_{\Lambda} = 0.7$, $h
= 0.7$ is adopted throughout this paper.

\section[]{The Survey Design}
\label{sec-design}

\subsection{Scientific goal}

The WiggleZ Dark Energy Survey is a major large-scale structure survey 
of intermediate-redshift UV-selected emission-line galaxies, focusing 
on the redshift range $0.2 < z < 1.0$.  The survey is designed to map a 
cosmic volume of about 1 Gpc$^3$, sufficient to measure the imprint of 
baryon oscillations in the clustering pattern at a significantly higher 
redshift than has been previously achieved.  The resulting measurements 
of cosmic distances and expansion rates, with uncertainties for each below $5\%$,
% {\it WJC: we need to be consistent right throughout this paper as to
% our accuracy/precision levels. We state less than 5\% here, having already
% stated in the previous section [core science goal (i)] that our goal is
% to measure the BAO scale to within 2\%!} 
will constitute a robust test of the predictions of the ``cosmological 
constant'' model of dark energy.  Detailed forecasts for the survey are 
presented in \citet{Blake2009a}.

\subsection{Choice of target galaxies}

The cosmological measurements from a large-scale structure survey
should be independent of the galaxy type used, given that the ``bias''
with which galaxies trace the underlying dark matter fluctuations is
expected to be a simple linear function on large scales
\citep{Coles1993,Scherrer1998}. In this sense, the choice of the
``tracer population'' of galaxies can be determined by observational
considerations such as telescope exposure times and the availability
of input imaging data for target selection.

The WiggleZ Survey targets UV-selected bright star-forming galaxies,
obtaining redshifts from the characteristic patterns of emission
lines.  A series of colour cuts, described below, are used to boost
the fraction of targets lying at high redshift.  The choice of blue
galaxies was motivated by several considerations: (1) emission-line
redshifts are generally obtainable in short 1-hour exposures at the
AAT, even in poor weather, with no requirement to detect the galaxy
continuum light; (2) the increasing star-formation rate density of the
Universe with redshift results in a significant population of
high-redshift $z > 0.5$ targets; (3) a large fraction of UV imaging
data for target selection was already available as part of the Medium
Imaging Survey conducted by the {Galaxy Evolution Explorer
satellite \citep[GALEX;][]{Martin2005}; (4) other large-scale structure surveys
such as the Sloan Digital Sky Survey (SDSS) and the forthcoming Baryon
Oscillation Spectroscopic Survey \citep[BOSS;][]{Schlegel2009} have}
preferred to observe Luminous Red Galaxies, allowing the WiggleZ
Survey to constrain any systematic effects arising from the choice of
tracer galaxy population; and (5) interesting secondary science is
achievable from studying bright star-forming galaxies at high
redshift.

The main disadvantage in the choice of WiggleZ Survey targets is that
emission-line galaxies possess a significantly lower clustering
amplitude than red galaxies, and hence larger galaxy number densities
are required to minimize the shot noise error in the clustering
measurement.  However, there is a benefit to the fact that WiggleZ
galaxies avoid the densest environments: non-linear growth of
structure---which erases linear-regime signatures in the power
spectrum such as baryon oscillations---becomes less significant in the
sample.

\subsection{Target numbers}

The WiggleZ Survey goal is to cover $1{,}000$ deg$^2$ of the equatorial 
sky in order to map a sufficiently large cosmic volume to measure the imprint of 
baryon oscillations in the clustering power spectrum.  The survey was 
designed such that the average galaxy number density $n_g$ is related to 
the amplitude of the galaxy clustering power spectrum $P_g$ on the 
relevant baryon oscillation scales by $n_g \sim 1/P_g$, implying that 
the contributions of sample variance and shot noise to the clustering 
error are equal.  This is the optimal survey strategy for a fixed number 
of galaxies (i.e.\ fixed observing time).

The galaxy selection cuts described below result in an average target
density of 340 deg$^{-2}$.  The overall survey redshift completeness,
allowing for repeat observations of objects originally observed in the
poorest conditions, is $70\%$.  {The survey goal is hence to measure
$\sim 340{,}000$ galaxies over $1{,}000$ deg$^2$, of which $\sim
240{,}000$ are expected to yield successful redshifts.}

We focus on the redshift range $0.2 < z < 1.0$ which we anticipate 
will yield our main baryon oscillation distance measurement.  {The 
fraction of redshifts lying in this range is 90 per cent,} 
therefore the average galaxy number density in the range $0.2 < z < 1.0$ 
is $n_g = 2.0 \times 10^{-4} h^3 \Mpc^{-3}$.  The matter power spectrum 
amplitude $P_m$ at the characteristic scale of baryon oscillations ($k = 
0.15 h \Mpc^{-1}$) is $3800 h^{-3} \Mpc^3$ at $z=0$ (for $\sigma_8 = 
0.9$). Assuming a galaxy bias of $b=1.21$ \citep{Blake2009a} and a 
growth factor $D=0.74$ (at $z=0.6$), the galaxy power spectrum amplitude 
at $z=0.6$ is $P_g = P_m D^2 b^2 = 3050 h^{-3} \Mpc^3$. Converting to 
redshift-space (using the boost factor $1 + \frac{2}{3} \beta + 
\frac{1}{5}\beta^2$ where $\beta=0.5$) we obtain $P_g = 4210 h^{-3} 
\Mpc^3$.  The survey value of $n_g \times P_g $ is therefore equal to 
0.84, close to the optimal condition of $n_g P_g = 1$.

{The average redshift completeness of an individual observation (or
``pointing'') is $60\%$ (we reach the value of 70\% per galaxy
mentioned above after repeating measurements made in poor conditions).
Thus the total number of individual spectra required is $N = 240{,}000
/ 0.6 = 400{,}000$.}  On average we can use 325 fibres per pointing and
obtain 7.4 pointings per night, and therefore can measure the required
$400{,}000$ spectra in 166 nights of clear weather.  This calculation
is summarized in Table~\ref{tab-design}.

\subsection{Survey fields}

A final constraint on the survey design is the availability of suitable 
optical imaging catalogues in regions of sky overlapping the GALEX UV 
data.  We require the optical data to provide more accurate positions 
for fibre spectroscopy and to refine our target selection.  The optical 
data we use are from the fourth data release of the Sloan Digital Sky 
Survey \citep[SDSS,][]{Adelman2006} in the NGP region and from the 
Canada-France-Hawaii Telescope (CFHT) Second Red-sequence Cluster Survey 
\citep[RCS2,][]{Yee2007} in the SGP region.

\begin{figure*}
\epsfig{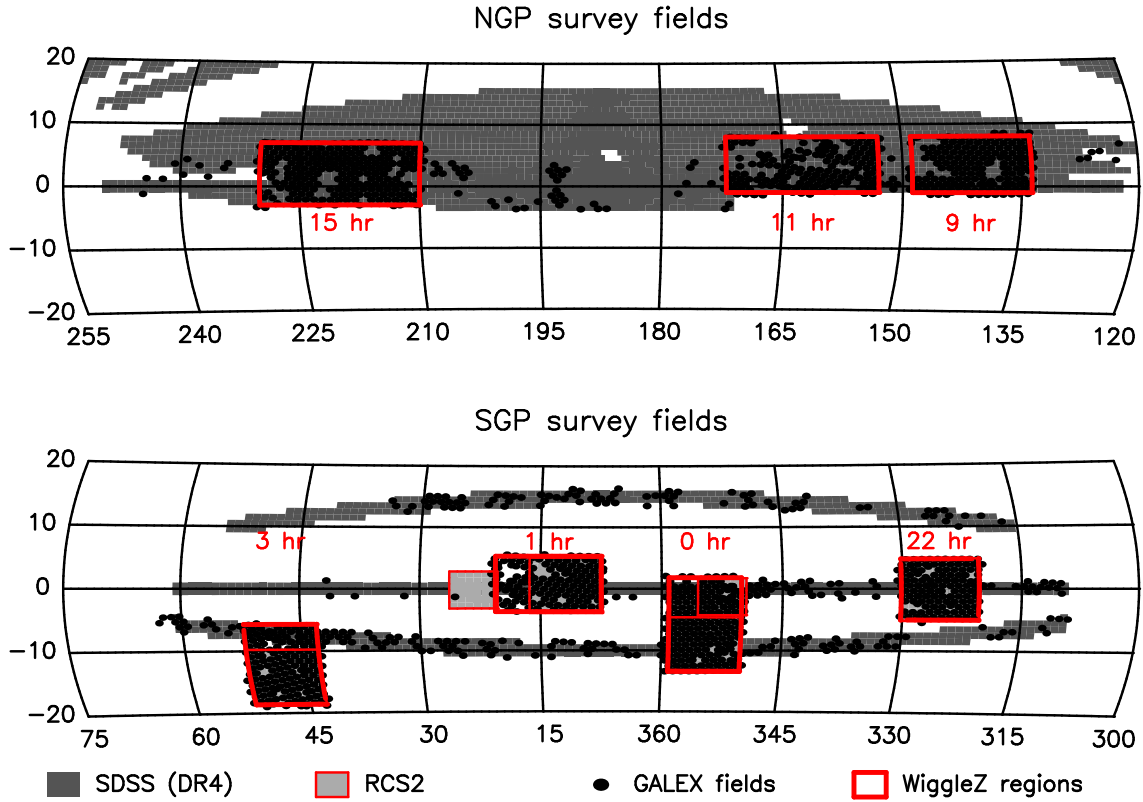}
\caption{The sky distribution of the seven WiggleZ survey
    regions compared to the coverage of the SDSS, RCS2 and GALEX
    data sets at the end of 2008.}
\label{figsurvey}
\end{figure*}

The WiggleZ survey area, illustrated in Figure~\ref{figsurvey}, is split 
into seven equatorial regions that sum to an area of 1000 deg$^2$, to 
facilitate year-round observing.  We require that each region should 
possess a minimum angular dimension of $\sim 10$ deg, corresponding to a 
spatial co-moving scale that exceeds by at least a factor of two the 
standard ruler preferred scale [which projects to (8.5, 4.6, 3.2, 2.6) 
deg at $z = (0.25, 0.5, 0.75, 1.0)$].  The boundaries of the regions are 
listed in Table~\ref{tab-survey}.  We have also defined high-priority 
sub-regions in each area (see Table~\ref{tab-surveyhigh}) which will be 
observed first to ensure that we have the maximum contiguous area 
surveyed should we be unable to observe the complete sample.

\begin{table}
\caption{Design of the Observational Programme}
\label{tab-design}
\begin{tabular}{lll}
\hline
Component     &  Value \\
\hline
Success rate (fraction $Q\ge 3$)  &   60\%  \\
%Efficiency   (fraction $z>0.5$)  &   70\%  \\
Target fibres/field  &   325   \\
Fields/ clear night  &  7.4   \\
%\hline
{\bf Rate (successful targets/night)} &   1442  \\
\hline
Required number of targets  &  240,000  \\
Rate per night (from above)         &   1442  \\
Required clear nights  &   166   \\
Weather factor      &   1.33  \\
{\bf Required total nights } &   221 \\
\hline
\end{tabular}
\end{table}

\begin{table}
\caption{Survey Regions: Full Extent}
\label{tab-survey}
\begin{tabular}{rrrrrr}
\hline
Name & RA$_{min}$ & RA$_{max}$ & Dec$_{min}$ & Dec$_{max}$ & Area\\
     & (deg)     & (deg)     & (deg)     & (deg)     & (deg$^2$) \\
\hline
  0-hr &  350.1 & 359.1  &  $-13.4$ & $+1.8$ &  135.7 \\
  1-hr &    7.5 &  20.6  &   $-3.7$ & $+5.3$ &  117.8 \\
  3-hr &   43.0 &  52.2  &  $-18.6$ & $-5.7$ &  115.8 \\
  9-hr &  133.7 & 148.8  &   $-1.0$ & $+8.0$ &  137.0 \\
 11-hr &  153.0 & 172.0  &   $-1.0$ & $+8.0$ &  170.5 \\
 15-hr &  210.0 & 230.0  &   $-3.0$ & $+7.0$ &  199.6 \\
 22-hr &  320.4 & 330.2  &   $-5.0$ & $+4.8$ &   95.9 \\
\hline
\end{tabular}
\end{table}

\begin{table}
\caption{Survey Regions: Priority Areas}
\label{tab-surveyhigh}
\begin{tabular}{rrrrr}
\hline
Name & RA$_{min}$ & RA$_{max}$ & Dec$_{min}$ & Dec$_{max}$ \\
     & (deg)     & (deg)     & (deg)     & (deg)     \\
\hline
  0-hr &  350.1  & 359.1  &  $-13.4$ &  $-4.4$    \\
  1-hr &    7.5  &  16.5  &   $-3.7$ &  $+5.3$    \\
  3-hr &   43.0  &  52.2  &  $-18.6$ &  $-9.6$    \\
  9-hr &  135.0  & 144.0  &   $-1.0$ &  $+8.0$    \\
 11-hr &  159.0  & 168.0  &   $-1.0$ &  $+8.0$    \\
 15-hr &  215.0  & 224.0  &   $-2.0$ &  $+7.0$    \\
 22-hr &  320.4  & 330.2  &   $-5.0$ &  $+4.8$    \\
\hline
\end{tabular}

Note: the high-priority region of the 22-hr field is the full field.
\end{table}

\section[]{Target Galaxy Sample}
\label{sec-target}

In this section we describe how we select our target galaxies from the
GALEX ultraviolet data combined with additional optical imaging
data. We stress that the target selection is motivated by our main
science objective of measuring the BAO scale. For this we need to
obtain as many high redshift galaxy redshifts as possible per hour of
AAT observing without necessarily obtaining a sample with simple
selection criteria in all the observing parameters. We nevertheless
discuss the selection process in detail in this section to facilitate
the use of the WiggleZ data for other projects that may need
well-defined (sub) samples.

\subsection{GALEX ultraviolet imaging data}
\label{sec-galexdata}

Our target galaxies are primarily selected using UV photometry from
{the GALEX satellite \citep{Martin2005}.}
This satellite is carrying out multiple imaging surveys in two
ultraviolet bands, in FUV from 135--175\,nm, and in NUV from 
175--275\,nm. Specifically, the WiggleZ survey uses the photometry of the GALEX
Medium Imaging Survey (MIS), which is an imaging survey of 1,000
square degrees of sky to a depth of 23\,mag in both bands. The
MIS consists of observations of 1.2\,degree diameter circular tiles,
with a nominal minimum exposure time of 1500\,seconds. Several hundred
extra GALEX observations will be dedicated to the WiggleZ
project. These are in addition to the original MIS survey plan, but
they are all taken with the standard MIS-depth exposures and cover
more contiguous sky areas due to improved bright-star avoidance
algorithms. Furthermore, GALEX observations in the original MIS plan
that overlap the WiggleZ survey regions are being observed at a higher
priority to maximise the fields available for the WiggleZ survey.

\begin{figure}
\epsfig{file=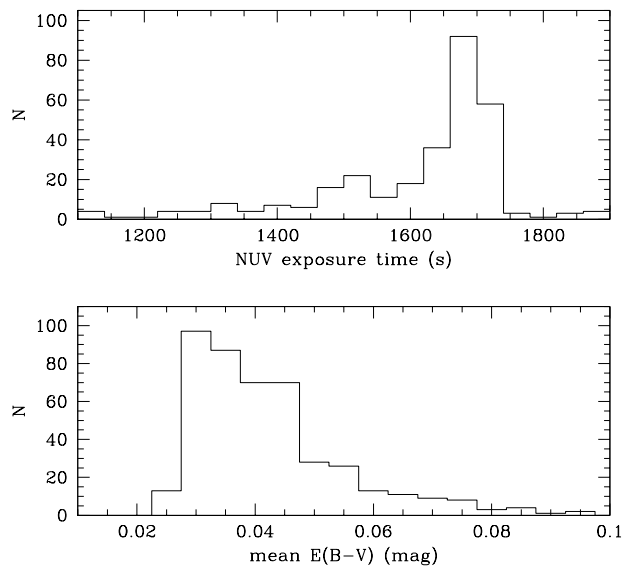,width=\linewidth,angle=0}
\caption{Properties of GALEX data used to select WiggleZ
  targets. Upper panel: distribution of exposure times. Lower panel:
  distribution of average dust extinction of each tile.}
\label{fig-tiles}
\end{figure}

The MIS data we use for WiggleZ have a range of exposure times, as
illustrated by the distribution shown in Figure~\ref{fig-tiles}. The
sharp peak at 1700\,seconds arises because this is the maximum possible
observing time in one satellite orbit. The figure also shows that
there is a range in average Galactic dust extinction across the
different GALEX observations. The effect of dust is discussed in
Section~\ref{sec-uniformity} below.

All GALEX observations for the WiggleZ project are processed by the
standard MIS data pipeline at Caltech to produce image catalogues
\citep{Morrissey2007}. The NUV images are processed in a standard way
with objects detected according to a threshold criterion using
SExtractor \citep{Bertin1996}. For the NUV detections we use the ``NUV
calibrated magnitude'' from the pipeline: this is the flux through an
elliptical aperture scaled to twice the Kron radius of each source
(termed ``MAG\_AUTO'' by SExtractor). The detection limit in the raw
NUV data is about 23 mag, as shown in Fig.~\ref{fig-uvcount} (a). We
used a different approach for the FUV photometry because relatively
few objects at the faint limit of our survey ($NUV\approx 22.8$) were
independently detected in the FUV band in the relatively short MIS
exposures. We therefore based the FUV photometry entirely on the NUV
detections. For each object in the NUV catalogue, the corresponding
FUV magnitude is obtained by measuring a fixed 12-arcsecond (8 pixel)
diameter circular aperture (``aperture 4'' in the pipeline) in the FUV
image at the NUV position. The full width half-maximum (FWHM) of the
GALEX point spread function (PSF) is 4.3 arc seconds in the NUV and
5.3 arc seconds in the FUV.  Our use of total NUV magnitudes and
aperture FUV magnitudes means that our $FUV-NUV>1$ colour selection
could, in principle, be biased by colour gradients in resolved
sources. Similarly we could suffer from source confusion with objects
separated in the NUV data still being close enough to contaminate the
FUV apertures, causing objects to incorrectly not satisfy the FUV-NUV
colour limit (see Section~\ref{sec-selection}).  In practice, the low
FUV signal means that very few sources are rejected by this colour
criterion.  The 1-$\sigma$ calibration uncertainties in the GALEX
photometry are 0.05 and 0.03 magnitudes in the FUV and NUV bands, 
respectively \citep{Morrissey2007}.

The astrometric precision of the MIS catalogue positions is 0.5 arc
second r.m.s., but this worsens with increasing distance from the
centre of a tile \citep{Morrissey2007}. The nominal FWHM of the PSF of
5 arc seconds also becomes poorer at increasing distance from the
centre of a tile. Additionally, uncertainties in the photometric
zero-points of the GALEX observations increase radially from the tile
centre. For these reasons, the WiggleZ survey only utilises the inner
1.1 degree diameter region of each GALEX tile, where the astrometry,
PSF and photometric zero-point uncertainties are negligibly different
to the nominal values.

At UV wavelengths, it is especially important to correct the
photometry for Galactic dust extinction. The standard GALEX pipeline
records the $E(B-V)$ extinction \citep[][hereafter SFD]{Schlegel98} 
for each detected source. We then apply the following corrections 
(including zero-point calibration to the AB system) to the NUV and FUV 
photometry
\citep{Wyder2007}:
\begin{eqnarray}
FUV &=& FUV_{raw} + 18.82 - 8.2 \times E(B-V) \nonumber \\
    & & +0.06\times E(B-V)^2,\\
NUV &=& NUV_{raw} + 18.82 - 8.2\times E(B-V) \nonumber \\
    & &  +0.67\times E(B-V)^2.
\end{eqnarray}
We note that the SFD dust maps have a resolution of 6.1\,arc min
and that the uncertainty in the reddening values is 16\%. 
We also note that SFD determined that it was not
reliable to calibrate the dust map normalisation from variations in
the number counts of optically-selected galaxies: this over-estimates
the dust amplitude because typical galaxy catalogues are both
magnitude and surface brightness limited. We further discuss the sensitivity
of our image catalogues to dust in
Section~\ref{sec-uniformity}.

\subsection{Optical Data}

In addition to the GALEX photometry, we use optical photometry from
the fourth data release of the Sloan Digital Sky Survey
\citep[SDSS,][]{Adelman2006} and from the CFHT Second Red-sequence
Cluster Survey \citep[RCS2,][]{Yee2007} to provide accurate astrometry
and improved target selection.

The SDSS photometry covers 7,000 square degrees of sky in 5 bands from
the near ultraviolet to the near infrared. The five bands are $u, g,
r, i$ and $z$, with magnitude limits (95 per cent point source
completeness) of 22.0, 22.2, 22.2, 21.3 and 20.5 mag respectively. The
median PSF FWHM (in $r$) is 1.4 arc seconds.  The photometric
calibration of the SDSS is nominally accurate to 0.02--0.03 magnitudes
\citep{Ivezic2004}, but the uncertainty at our faint detection limit
of $r=22.5$ is more likely to be 0.2--0.3 magnitudes. We use the SDSS
``model'' magnitudes, calculated on the AB system. The location of
objects detected in these bands is known to an rms astrometric
accuracy of 0.1 arc seconds.

The RCS2 is an imaging survey of 1,000 square degrees of sky in 3 of the SDSS
bands: $g, r$ and $z$. The magnitude limits (5-sigma point source
limits) are 25.3, 24.8, and 22.5 respectively, significantly fainter
than in the SDSS.  The typical seeing in the RCS2 imaging is ($0.6\pm
0.1$) arc seconds for the best half of the data and ($0.8\pm 0.1$) arc
seconds for the remainder.  The internal photometric precision of the
RCS2 is close to that of the CFHT Legacy Survey \citep{Ilbert2006},
i.e.\ 0.04 mag in each band; the magnitudes are on the AB system,
calculated from a curve-of-growth analysis. The colours are measured
in smaller (adaptive) apertures and then scaled to the total magnitude
of the source in the band with the strongest detection (usually $r$).
The RCS2 has an astrometric accuracy of 0.15 arc seconds, comparable
to the astrometric precision of SDSS.

The precise astrometry of both these optical data sets is crucial to
the WiggleZ survey, because the combination of the
point-spread-function and astrometry of the GALEX source detections is
too poor for the optical fibres used to feed light into the AAOmega
spectrographs, which subtend an angle of 2 arc seconds on the sky.  In
all our processing we use dust-corrected optical catalogues. For SDSS,
the dust correction is provided as part of the public data and for
RCS2 we apply a dust correction ourselves: in both cases this is the
standard SFD correction.

\subsection{Matching UV to optical samples}

The WiggleZ targets are selected from the GALEX UV photometry and then
combined with either SDSS or RCS2 optical photometry. In the NGP
region we combine the GALEX photometry with SDSS photometry, and in
the SGP region we combine it with RCS2 photometry. The respective
catalogues are combined by selecting the closest optical match to each
GALEX source within a radius of 2.5 arc seconds. For the surface densities of
the three data sets, this corresponds to 95 per cent confidence in a
GALEX-SDSS match and 90 per cent confidence in a GALEX-RCS2
match. (The lower confidence for the RCS2 matches is mainly a
consequence of the higher number density of sources in the RCS2 data.)
We use the optical position for each matched source in our target
catalogues as these are more precise than the GALEX positions.

Once an object is matched to the optical data, we can then apply
further colour limits to refine our colour selection (see
Section~\ref{sec-selection}) noting that the $NUV-r$ colour is a total
colour. However, the large size of the GALEX PSF (FWHM 4.3 arc seconds
in the NUV) compared to the optical imaging means that the colours may
be distorted if multiple objects are merged in the GALEX photometry,
but are resolved in the optical imaging. In this case the GALEX
photometry would be too bright compared to the optical measurement. As
discussed in Section~\ref{sec-properties}, this is possible given the
evidence for multiple sources we present in Fig.~\ref{fig-morphology}.

\subsection{Uniformity of GALEX Data}
\label{sec-uniformity}

\begin{figure}
\epsfig{file=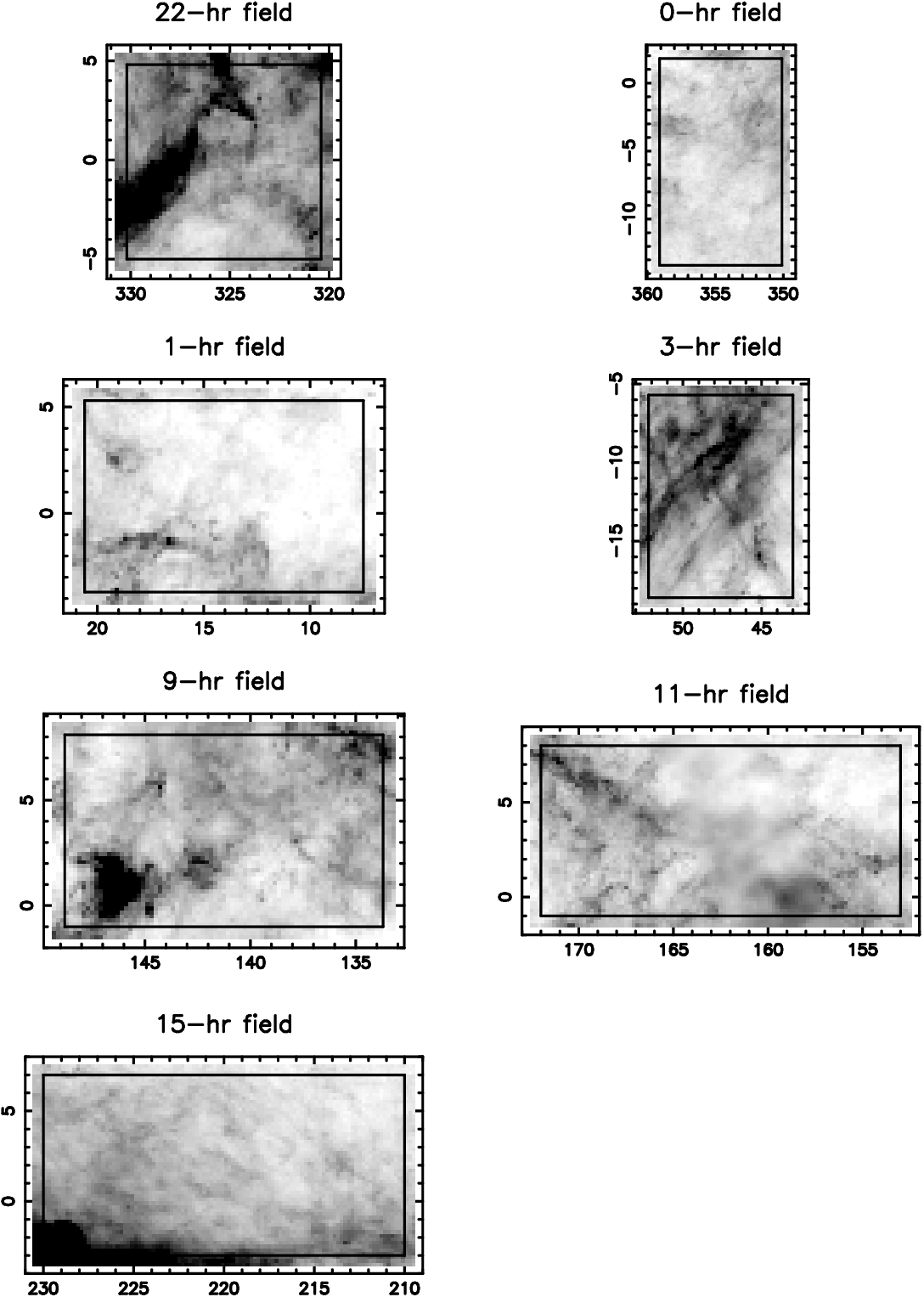,width=1.0\linewidth,angle=0}
\caption{Distribution of dust in each of the survey fields. The grey
  scale shows the \citet{Schlegel98} extinction values such that 
white is $E(B-V)=0.02$ and black is $E(B-V)=0.1$. The horizontal and
vertical axes in each panel denote the Right Ascension and
Declination, respectively, measured in degrees.
}
\label{fig-dust}
\end{figure}

An important requirement for the survey is that no artificial
structure be introduced by variations in the input catalogues. This is
particularly true for the GALEX data as the tile diameter (1.1
degrees) is close to the BAO scale (see Section \ref{sec-design}). The
primary source of non-uniformity is likely to be foreground Galactic
dust, although we also test the data for any calibration offsets. To
demonstrate the importance of dust, we show the range of {\em average}
dust extinction across each tile, for a set of GALEX tiles in
Fig.~\ref{fig-tiles} (lower panel). This range ($0.03\lesssim E(B-V)
\lesssim 0.07$; see Section \ref{sec-galexdata} for the conversion to
$A_{NUV,FUV}$) is sufficient to affect the detection rates as we show
below.  In Fig.~\ref{fig-dust} we present images of the
SFD dust maps in each field. These clearly show
structure on scales similar to and smaller than the BAO scale, so it
is very important to quantify the effect of dust on our input
catalogues and correct for it if necessary. We also note an additional
effect: some of the regions of higher dust are associated with
Galactic nebular emission. One such example is the Eridanus loop which
coincides with the diagonal dust lane visible in the 3-hour field in
Fig.~\ref{fig-dust}. Spectra of targets in such regions can be
contaminated by rest-wavelength nebular emission lines, confusing the
target redshift measurement.

We tested the GALEX photometry in two ways: first by analysing the
detected number counts and secondly by comparing the photometry of
objects measured in overlapping GALEX tiles. The first test addresses
all factors affecting image detection (notably dust) and the second
tests the internal calibration of the photometry. The second tests did
not reveal any evidence for significant differences in absolute
calibration between overlapping tiles, so these will not be discussed
in any more detail (small offsets were measured, but these were
uncorrelated the numbers of images detected on the respective tiles).

\begin{figure*}
\begin{minipage}{1.05\linewidth}
\hspace*{-4mm}
\epsfig{file=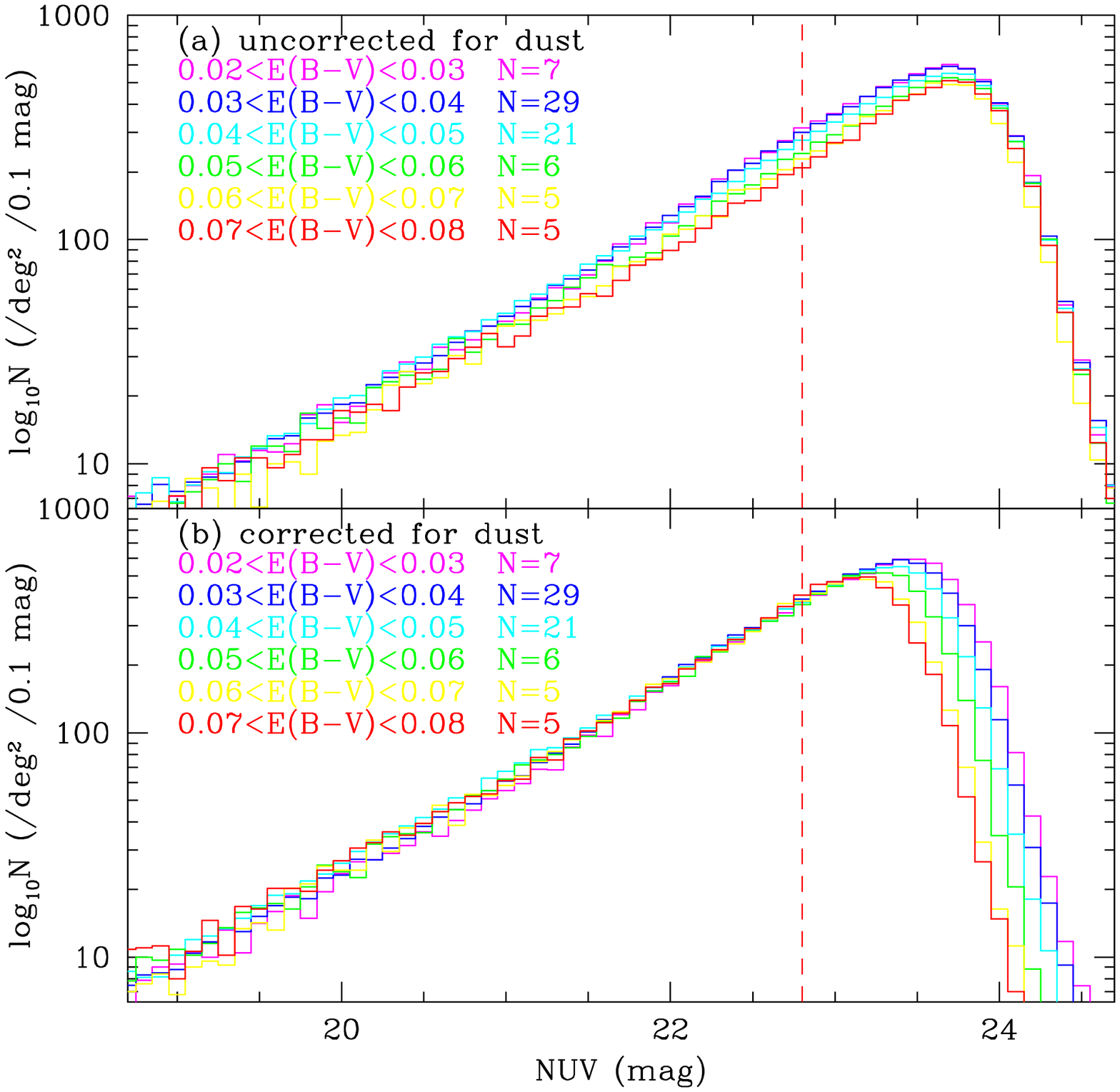,width=0.52\linewidth,angle=0}
\hspace*{-6mm}
\epsfig{file=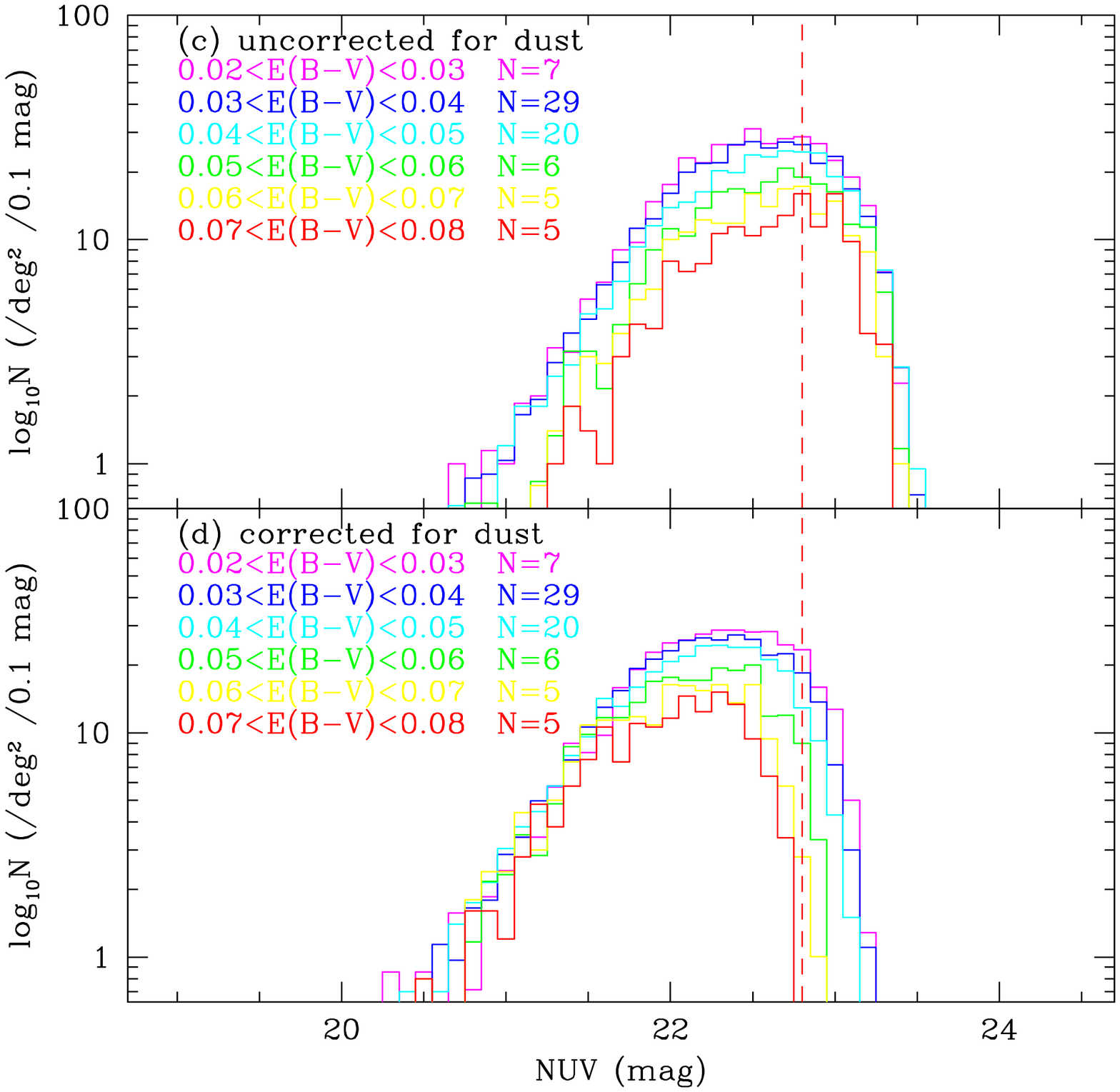,width=0.52\linewidth,angle=0}
\end{minipage}

\caption{Number counts of GALEX NUV detections in tiles from the
  15-hour region as a function of dust extinction. The NUV magnitude
  limit of the WiggleZ survey is indicated by a vertical dashed line
  in each plot. In the left hand panels (a, b) we plot all objects
  detected before and after the dust correction. The counts are binned
  by the average dust extinction in each tile, one curve for each bin
  (see key). The upper panel shows the raw counts; in the lower panel
  the dust correction has been applied to every individual object. The
  improved agreement of the power-law region of the number counts in
  panel b demonstrates the effectiveness of the dust correction. The
  right hand panels (c, d) are similar, but only count objects that
  satisfy the WiggleZ target selection criteria. Panel d shows that
  the target numbers at the NUV survey limit are more sensitive to
  dust than the total counts of detections shown in panel b.  }
\label{fig-uvcount}
\end{figure*}

In Figure~\ref{fig-uvcount}(a,b) we show the differential number counts
of all sources detected in the NUV tiles in the 15-hour region. In the
upper panel the counts are plotted as a function of raw observed
magnitude and in the lower plot, as a function of dust-corrected
magnitude. In each case the tiles are grouped according to the average
dust extinction for each tile and the lines plotted give the average
counts for each group of tiles as indicated by the key. We further
limited the choice of tiles to those with exposures in the range
1600--1700 seconds to focus on the role of extinction. Our first
observation from the plots is that the dust correction is working as
expected: the uncorrected counts all show the same faint cut-off in
raw magnitude, but they have different normalisations of their
power-law slopes. In contrast, the dust-corrected counts show the same
power-law normalisation, as desired.

Although we have shown that calibration variations in the GALEX data
are minimal, the faint magnitude cut off in the counts has a strong
effect on the number of WiggleZ targets selected as shown in
Fig.~\ref{fig-uvcount}(c,d). Here we show the number counts of WiggleZ
targets, having applied the survey selection criteria (see
Sec.~\ref{sec-selection}) to the raw counts. This shows that, unlike
the raw counts, the WiggleZ target counts at our survey limit
($NUV<22.8$) are strongly affected by dust.  The mean counts per tile
vary by a factor of two over the full range of extinction, although 70
per cent of the tiles lie in a narrow range of extinction
($0.025<E(B-V)<0.045$) for which the counts vary by only 10 per
cent. For the WiggleZ clustering measurements we will deal with this
effect by scaling the random catalogues to follow the GALEX magnitude
limits. We have tested this approach by measuring the angular power
spectrum of the galaxies and comparing that to theoretical
models. After scaling our selection function to allow for the combined
effects of dust and exposure time on the GALEX counts we find there is
no evidence for any large-scale systematic variation
\citep{Blake2009a}.  For measurements of the luminosity function of
the WiggleZ sources, however, the correction needs to be with respect
to the underlying power law and is therefore much larger. This will be
discussed in more detail in later papers.

\subsection{Galaxy Selection}
\label{sec-selection}

\begin{figure}
\epsfig{file=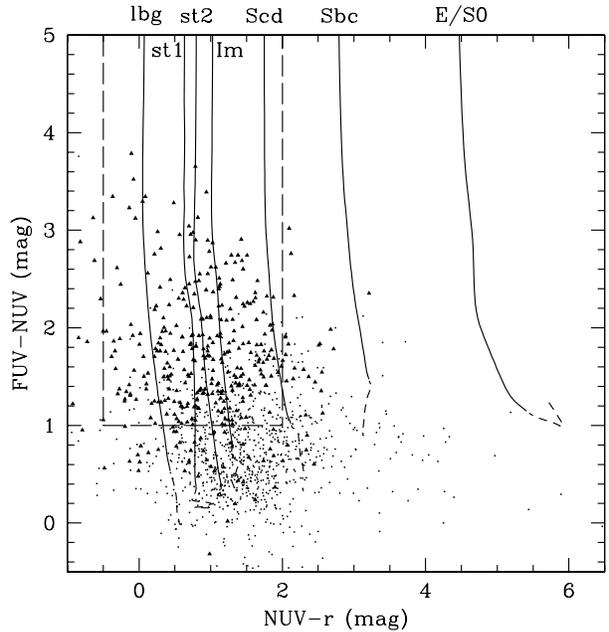,width=\figwidth\linewidth,angle=0}

\caption{The basic WiggleZ target selection criteria. The rectangle
  marked by the dashed line denotes the region in FUV-NUV,NUV-r colour
  space where WiggleZ targets lie. The solid lines are the model
  tracks for seven galaxy templates as a function of redshift for $z >
  0.5$.  The first track, from left to right, is for a Lyman break
  galaxy template \citep[Figure 1 in][]{Steidel1996}. The remaining
  tracks are two
  starburst templates \citep{Kinney1996} and a series of (Im to E/S0)
  local galaxy templates \citep{Coleman1980}, all extended in
  wavelength range by matching them with \citet{Bruzual1993} models
  (Hsaio-Wen Chen, private communication).  The dashed lines at the
  end of the solid lines are these models for $z \le 0.5$. Also shown
  as points are galaxies from deep GALEX imaging that were detected in
  all three bands and have published redshifts from the DEEP2 survey
  \citep{Davis2003}. The triangles indicate objects with redshift $z >
  0.5$.}

\label{basicTselect}
\end{figure}

The main criteria used to select the WiggleZ survey targets from the
GALEX and optical photometry are a $FUV-NUV$ colour cut to select
high-redshift galaxies, and a $NUV-r$ colour cut to select emission
line galaxies. These are shown in Fig.~\ref{basicTselect}, a
colour-colour plot of $FUV-NUV$ vs.\ $NUV-r$. In the diagram we show
the WiggleZ selection limits (dashed rectangle), data points from
combined DEEP2 \citep{Davis2003} and GALEX observations, and model tracks for several
galaxy types over a range of redshifts.

The UV colour selection (specifically $FUV - NUV \geq 1$ or $FUV$
undetected: a ``drop-out'') selects galaxies at redshifts $z\geq 0.5$
because the Lyman break enters the FUV filter at a redshift of $z
\approx 0.5$. We note that in the MIS UV data used in our survey, for
fainter galaxies (near our $NUV$ detection limit), the $FUV$
measurements have very low signal-to-noise, so the drop-out selection
criterion is very crude. (The data points in Fig.~\ref{basicTselect}, 
by contrast, were measured in much deeper GALEX data than used for our
survey.)

The diagram also shows how we use the $NUV-r$ colour. The $NUV-r \leq
2$ limit selects bluer star-forming galaxies, and the $-0.5 \leq NUV-r
$ limit helps remove spurious matches between the GALEX and optical
data sets. These two colour cuts were chosen based on the colours of
galaxy models as a function of redshift. The colours of these galaxy
models and the colour cuts used for the WiggleZ target selection are
shown in Fig.~\ref{basicTselect}.

For the GALEX NUV photometry we impose a faint magnitude limit of 22.8
in NUV.  In addition, we also require that the NUV flux measurement
has a signal-to-noise $\geq 3$ to minimise the number of false
detections. This requirement becomes increasingly important for
dustier regions of the sky: in regions of low dust ($E(B-V)\approx
0.03$) the signal-to-noise cut removes about 10 per cent of the raw
$NUV \leq 22.8$ sources, but in high dust regions ($E(B-V)\approx
0.06$) it removes 25 per cent of the sources.

For the optical photometry (both SDSS and RCS2) we impose a magnitude
range of $20<r<22.5$. The faint limit is set by the SDSS survey
detection limit; the bright limit is to avoid low-redshift
galaxies. As a consequence of this bright $r$ magnitude limit, the
$NUV-r$ colour cut results in a bright magnitude limit in the NUV of
19.5, although this is not explicitly part of the WiggleZ target
selection criteria. A further consequence of the $NUV-r$ colour cut
(and indeed the average $NUV-r$ colours) is that the optical
magnitudes of the selected galaxies do not peak at the faint $r=22.5$
limit, but instead peaks at $r\approx 21.5$, about a magnitude
brighter. Note that we do not apply any morphological selection to
remove objects classified as ``stellar'' in the optical imaging
because virtually no stars satisfy our photometric selection criteria.
These basic selection criteria are listed in the first
section of Table~\ref{tab-selection}.

\begin{figure}
\epsfig{file=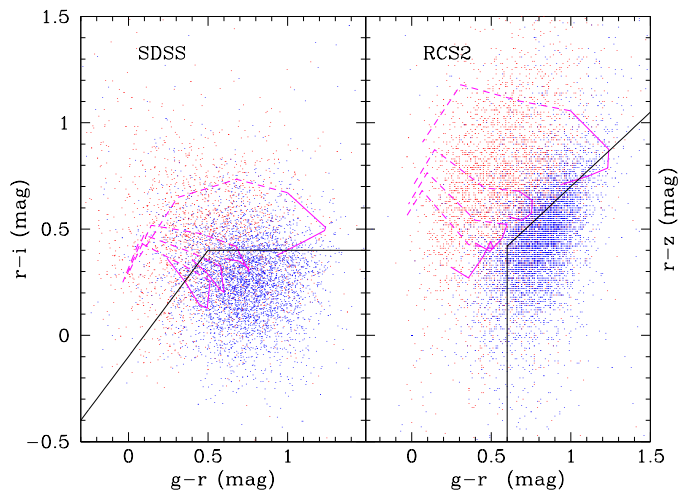,width=0.8\linewidth,angle=270}
\caption{The selection criteria applied to avoid low-redshift targets
  in the two survey regions. The points in each diagram are WiggleZ
  galaxies observed before we introduced the low redshift rejection
  cuts. Galaxies with redshifts below $z=0.6$ are plotted as blue
  points; the rest are plotted in red. Galaxies below the lines in
  each panel are rejected from the target catalogue as they have a
  high probability of being at low redshifts. Also shown as magenta
  lines are PEGASE-2
  \citep{Fioc1997} evolutionary tracks of four galaxy models ranging
  from E/SO (upper) to Irr (lower) types. Each model ranges from $z=0$
  to $z=1$ with the high redshift ($z>0.5$) section dashed.
}
\label{fig-karl}
\end{figure}

\begin{figure}
\epsfig{file=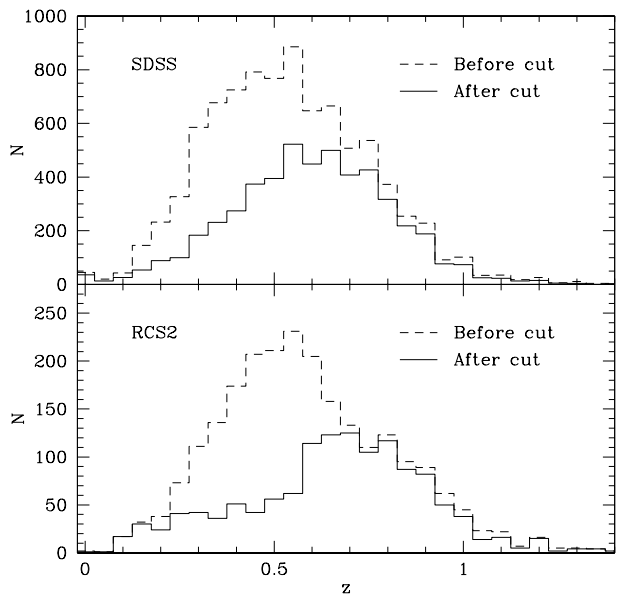,width=\figwidth\linewidth,angle=0}
\caption{The result of the optical colour cuts used to reduce the
  number of low-redshift targets. The top panel shows the effect in
  the Northern (SDSS) regions of the survey and the lower panel shows
  the Southern (RCS2) regions. In each panel the dashed histogram
  shows the redshift distribution of all galaxies observed before the
  dates on which the colour cuts were applied in each section of the
  survey (see dates in Table~\ref{tab-dates}). In each panel the
  solid histogram shows the results of retrospectively applying the
  colour cuts to these same objects.  In both regions the cuts result
  in a significantly higher median redshift; we note that the deeper
  photometry in the RCS2 regions allows a more effective cut to be
  applied in the lower panel. }
\label{fig-selection}
\end{figure}

A further selection based on optical colour is used to reduce the
number of targets selected with redshifts below $z=0.5$. These arise
because the $FUV-NUV$ selection is imperfect due to photometric errors
--- many of our targets are at the limits of the GALEX photometry in
typical MIS exposures.  We do this by rejecting galaxies whose optical
colours and magnitudes indicate they are very likely to be lying at
lower redshifts (as shown in Fig~\ref{fig-karl}).  Two different sets
of low redshift rejection (LRR) criteria are used in the SDSS and RCS2
regions respectively, as listed in Table~\ref{tab-selection}. In the
SDSS data we are near the faint limit of the survey, so can only apply
this to the brighter galaxies. However the brighter galaxies are more
likely to be at lower redshift so this ameliorates this disadvantage.
We therefore start by applying $g$ and $i$ magnitude limits in order
to secure accurate colours.  Then we apply a ($g-r, r-i$) colour cut
to select the 400 nm break in galaxies at redshifts $z<0.5$. In the
RCS2 regions the optical data go much deeper, so no magnitude limits
are required, but there are no $i$ data, so we have to use a $(g-r,
r-z)$ colour cut which is less precise at discriminating at $z<0.5$.
The success of these selection criteria is demonstrated in
Fig.~\ref{fig-selection} where we show redshift distributions of the
objects observed with and without the LRR colour cuts.

\begin{table}
\caption{Photometric selection criteria for WiggleZ galaxies}
\label{tab-selection}
\begin{tabular}{ll}
\hline
Criterion  &  Values     \\
\hline
\multicolumn{2}{l}{\em Select targets satisfying all these basic criteria:}\\ 
Magnitude &  $NUV<22.8$    \\
Magnitude &   $20 < r < 22.5$ \\
Colour    &  $(FUV-NUV) > 1$ or no FUV  \\
Colour    & $-0.5 < (NUV-r) < 2$  \\
Signal    &  $S/N_{NUV}>3$ \\
Optical Position  & matches within 2.5 arc seconds \\
\hline
\multicolumn{2}{l}{\em Then reject targets satisfying these:}\\ 
LRR$_{SDSS}$ & $g < 22.5$, $i < 21.5,$ \\
  & $(r-i) < (g-r-0.1),  (r-i) < 0.4$ \\
LRR$_{RCS2}$ & $(g-r) > 0.6, 
(r-z) < 0.7(g-r)$ \\
\hline
\end{tabular}
\end{table}

\begin{figure}
\epsfig{file=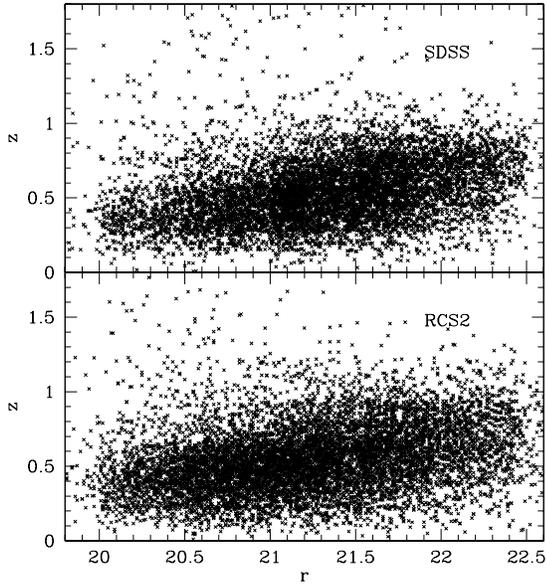,width=\figwidth\linewidth,angle=0}
\caption{The correlation between redshift and $r$-band flux for the
  observed WiggleZ galaxies. As there is a weak correlation in both
  the SDSS (top panel) and RCS2 (lower panel) samples, we observe
  targets with fainter $r$ magnitudes at a higher priority in order to
  increase the median redshift of the sample.
  }
\label{fig-fluxredshift}
\end{figure}

We also apply a prioritisation scheme to our target allocation as
given in Table~\ref{tab-priority}. The priorities given in the table
are used when observing so that higher-priority objects are observed
first. This is done for two reasons. First, there is a weak
correlation between $r$ magnitude and redshift (shown in
Fig.~\ref{fig-fluxredshift}) so this also serves to select
high-redshift objects. Secondly, this approach means that our later
observations will be of brighter objects, allowing us more flexibility
in the final stages of the observational campaign.

\begin{table}
\caption{Priority selection scheme for WiggleZ observations}
\label{tab-priority}
\begin{tabular}{ll}
\hline
Criterion  &  Priority   \\
\hline
repeat observations & 9 (highest) \\
white dwarf calibrators & 9 \\
{\em New targets$^1$  by magnitude:} \\
$22.0 < r < 22.5$ & 8     \\
$21.5 < r < 22.0$ & 7     \\
$21.0 < r < 21.5$ & 6     \\
$20.5 < r < 21.0$ & 5     \\
$20.0 < r < 20.5$ & 4     \\
{\em Other objects:} \\
Additional projects & 3 \\
Old good weather targets ($Q=$1,2) & 2 \\
Old good weather targets ($Q=$3--5) & 1 (lowest) \\
\hline
\end{tabular}

Note 1: The ``new'' targets include objects that were previously
observed in bad weather without obtaining a redshift.
\end{table}

Before observing we make a final check of all the target objects by
visually inspecting them on SuperCOSMOS \citep{Hambly2001} sky survey
images to ensure none are artifacts associated with extremely bright
($r<12$) Galactic stars. This happens occasionally with automated data
catalogues because bright objects can be incorrectly segmented into
smaller, fainter objects, which are then matched to GALEX NUV
detections. This is a problem, because the signal from these bright
objects swamps adjacent spectra in such a way that we lose as many as
30 spectra from an observation.

The final result of the whole selection process is a target catalogue
with an average surface density of 350 deg$^{-2}$ which is well
matched to our survey goals. The WiggleZ target galaxies represent a
small fraction of optical galaxies. Analysis of a sample of DEEP2
\citep{Davis2003} galaxies for which deep optical and GALEX
observations are available shows that in the SDSS regions, WiggleZ
selects $2.2\pm0.2$ per cent of the ($r<22.5$) optical galaxies. In
the RCS2 regions, WiggleZ selects $2.6\pm 0.2$ per cent of these same
galaxies. These low percentages are mainly due to the WiggleZ NUV flux
limit and NUV$-r$ colour cuts which only select galaxies with very
high current star formation rates, but we also note that the median
$r$ magnitude of the WiggleZ galaxies is brighter than $r=22.5$ (see
Fig.~\ref{fig-fluxredshift}).

\section[]{Spectroscopic Observations}
\label{sec-spectroscopy}

The major component of this project is our large, 220-night observing
campaign on the AAT. In this section we describe the management of
this campaign, the procedures required to configure our AAOmega
observations and the data processing.

\subsection{AAOmega Spectrograph and Observing Setup}

The efficiency of the WiggleZ redshift survey is critically dependent
on the high sensitivity and multiplexing gain made possible by the new
AAOmega spectrograph and existing 2dF fibre positioner on the AAT
\citep{Smith2004,Saunders2004}. AAOmega is a bench-mounted, fibre-fed
spectrograph consisting of blue and red arms split by a
dichroic. Volume phase holographic (VPH) gratings are utilized
resulting in an improved efficiency over reflection gratings. The
bench for the spectrograph is mounted in a mechanically and thermally
stable coud\'e room. For more details about the performance
characteristics of AAOmega, see \citet{Sharp2006}.

In multi-object mode the combination of the 2dF positioner and AAOmega
spectrograph allows observations with a total of 392 fibres over a
circular field of view with a diameter of two degrees. The angular
size of each fibre on the sky is 2 arc seconds. Our survey utilizes
the 580V and 385R gratings in the blue and red arms, respectively,
both providing a resolution of $R \approx 1300$. This gives a
dispersion of about 0.11 nm pixel$^{-1}$ in the blue arm and 0.16 nm
pixel$^{-1}$ in the red arm.

The observable wavelength range of the system is 370--950 nm. The blue
limit is determined by the fibre transmission characteristics (over
the 38m-long fibres) and the red limit is due to the CCD response. In
our early observations, we used the standard dichroic with the cut
over centred at 570 nm. As we also required a small spectral overlap
between the blue and red arms, this fixed the low-wavelength limit of
the red spectrograph which could then only cover up to a maximum
wavelength of 850 nm.  We alleviated this limit on observing redder
wavelengths by purchasing a new dichroic for AAOmega, in partnership
with the AAO, to extend our wavelength coverage.

The original AAOmega dichroic beam splitter cuts over at 570 nm while
the new dichroic has a cut over at 670 nm. The new dichroic was
specially designed for our project to allow improved spectral coverage
out to the system limit of 950 nm, using the 385R grating, while still
allowing continuous spectral coverage down to a new blue limit of
470nm (see Table~\ref{tab-aaomega}). This redder wavelength range
enables more reliable redshift identification of the emission-line
galaxies in our sample. Specifically, an increase of 100 nm at the red
end allows the detection of secondary emission-lines such as
H$\beta$/[OIII] lines at higher redshifts up to $z\approx 0.95$. At
redshifts above $z\approx 0.95$ we often rely on just the [OII]
(372.7, 372.9 nm) doublet for redshift identification, but we are
aided by the fact that the doublet is often resolved at these
redshifts. The new dichroic was first used for our 2007 August
observing run (see Table~\ref{tab-dates}). About 25 per cent of the
100,138 reliable redshifts measured up to the end of Semester 2008A on
2008 May 15 were taken with the original dichroic; the rest used the
new dichroic which will be used for the remainder of the survey.

% OLD: 19/2/06 -- 5/8/07 All spectra N = 61555
% NEW: 6/8/07 -- 15/5/08 All spectra N = 24137
% NEW: 6/8/07 -- 31/12/08 All spectra N = 311113
% both 19/2/06 -- 31/12/08 All spectra N = 372668
%
% Good galaxy redshifts z>0.0025
% 31/12/08  all       N  =  118471
% 31/12/08  0.5<z<1.5 N  =   70167
% 31/12/08  0.5<z     N  =   76735
%
% NED comparison

\begin{table}
\caption{The two AAOmega setups used for our observations}
\label{tab-aaomega}
\begin{tabular}{lrrrr}
\hline
Dichroic  &  $\lambda_{C,Blue}$ & Blue Range  &  $\lambda_{C,Red}$ &
Red Range \\
\hline
570 nm (old) &  477  &  370--580 & 725 & 560--850 \\
670 nm (new) &  575  &  470--680 & 815 & 650--950 \\
\hline
\end{tabular}

Note: for each dichroic the table lists the central wavelength setting
and observable wavelength range for the blue and red arms of the
spectrograph. For each VPH grating the blaze wavelength was set to be
the same as these central wavelengths for maximum efficiency. All units are nm.
\end{table}

\begin{table}
\caption{Significant Dates in our Observing Campaign}
\label{tab-dates}
\begin{tabular}{llr}
\hline
Date  &  Activity & $N_z$\\
\hline
2006 February & Pilot observations \\
2006 August 19 & Start of survey (semester 2006B) \\
2007 April 11 & Start of using LRR in SDSS fields&  12445\\
               & Start of 0.7 deg radius observing& \\
2007 June 8--21 & Observations without blue camera& 24740 \\
2007 August 7 & Start of new dichroic use & 26293 \\
2007 August 14 & End of 0.7 deg radius observing& 38540\\
2007 October 4 & Start of using LRR in RCS2 fields& 42179 \\
2008 May 13 & End of Semester 2008A and \\
& half-way point after 112 nights. & 100138\\
%2008 Dec 13 & & 119253 \\
\hline
\end{tabular}

Note: $N_z$ is the running total number of good redshifts ($Q\ge3$)
measured on each date.
\end{table}

\subsection{Blank Sky and Guide Star Positions}

The 2dF/AAOmega system, like other multi-object fibre spectrographs
requires fiducial (``guide'') stars to align the field accurately and
blank sky positions where a subset of the target fibres can be placed
to sample the sky background.

\subsubsection{Selection of Blank Sky Positions}

The blank sky positions must satisfy two criteria: their number
density on the sky should be high enough to ensure we can always
configure at least 25 sky positions in any given 2dF observation, and
each must be sufficiently far from any other object on the sky to
avoid contamination. We chose a density of 40 sky positions per square
degree, corresponding to 120 per 2dF observation. This was judged to
be sufficient to allow 25 sky fibres to be configured after all the
target and guide fibres were positioned.

We selected the sky positions automatically from the image catalogues
for each region by setting down a square grid of the desired surface
density. Any grid positions too close to a catalogue image were moved
(randomly) within the corresponding grid cell as necessary until they
satisfied a proximity test.  The proximity test for neighbouring
objects was a function of their apparent $r$ (or $R$) magnitude: the
sky position was moved if there were neighbours satisfying $r<12.5$, 
$r<16$, $r<17.5$, or $r\ge 17.5$, 
within distances of 1.5, 1, 0.5, or 0.25 arc
minutes respectively.

\subsubsection{Selection of Guide Star Positions}

The guide stars for AAOmega must satisfy strict photometric and
astrometric criteria. Most importantly, their positions must be
accurate and on the same reference frame as the target objects. The
new 2dF/AAOmega system has 8 fibre guide bundles on each plate so, in
principle, observations are still possible if one or two of the guide
stars are not suitable. However the diagnosis of the defective stars
slows down the observing process, to the extent that they would amount
to a significant time loss over the course of a large survey like WiggleZ.

To satisfy the astrometric criterion we selected guide stars from
different sources in each of the two survey regions. In the northern
SDSS survey regions we selected guide stars directly from the same
SDSS catalogues as we used for our optical target photometry. We
used the SDSS automated image classification flags to select stars
from the catalogues (``mode''=1, ``type''=6, ``probPSF''=1).

In the southern (RCS2) regions we were unable to obtain guide star
positions from our imaging data as they are much deeper and hence most guide
stars are saturated, so they have unreliable positions. We therefore
chose guide stars for these regions directly from the USNO-B
astrometric catalogue \citep{Monet2003} which was used to provide the
astrometry for the RCS2 data (with fainter USNO-B stars that were not
saturated in the RCS2 images). We undertook extensive testing of the
USNO-B guide star astrometry, comparing the positions to both SDSS
data (in the limited regions where they overlap) and also with the
all-sky 2-MASS catalogue data \citep{Skrutskie2006}. In the regions
tested we found small but significant systematic position differences
between 2-MASS and USNO-B; for the SDSS versus 2-MASS comparison, we 
found no systematic difference. In all cases the r.m.s.\ scatter in the 
position differences (less than 0.2 arc seconds in each coordinate) was 
consistent with the published uncertainties in the respective catalogues. 
The offsets are tabulated in Table~\ref{tab-offset}.  We have not adjusted 
the RCS2 or USNO-B
astrometry for the small systematic offsets with respect to the
SDSS/2-MASS astrometry, as this is not necessary for our spectroscopic
observations.  There is no measurable offset between the astrometry of
the RCS2 data (our targets) and the USNO-B catalogue (our guide
stars), so AAOmega observations using these guide stars will ensure
the target fibres are correctly positioned.

% Offsets RCS-2MASS
% 00hr N=5510
% # mean diff: ra=    -0.112, dec=     0.117, total=    0.166
% # rms  diff: ra=     0.243, dec=     0.244, total=    0.344
% # SEM                0.003           0.003
% 
% 01hr N=6293
% # mean diff: ra=    -0.133, dec=     0.102, total=    0.168
% # rms  diff: ra=     0.301, dec=     0.271, total=    0.405
% # SEM                0.004           0.003
% 03hr N=5308
% # mean diff: ra=    -0.141, dec=    -0.026, total=    0.143
% # rms  diff: ra=     0.178, dec=     0.166, total=    0.243
% # SEM                0.002           0.002
% 22hr  N=8604
% #  mean diff: ra=    -0.031, dec=     0.137, total=    0.141
% #  rms  diff: ra=     0.287, dec=     0.272, total=    0.395
% #  SEM                0.003           0.003

\begin{table}
\caption{Astrometry Offsets in RCS2 Regions}
\label{tab-offset}
\begin{tabular}{rrr}
\hline
Name & RA$_{USNO}-$RA$_{2MASS}$ & Dec$_{USNO}-$Dec$_{2MASS}$ \\
     & (arc seconds)     & (arc seconds)  \\
\hline
  0-hr &  $-0.112\pm 0.003$ &  $0.117\pm 0.003$  \\
  1-hr &  $-0.133\pm 0.004$ &  $0.102\pm 0.003$  \\
  3-hr &  $-0.141\pm 0.002$ & $-0.026\pm 0.002$  \\
 22-hr &  $-0.031\pm 0.003$ &  $0.137\pm 0.003$  \\
\hline
\end{tabular}
\end{table}

We did find that the USNO-B guide star positions, when compared to
2MASS, displayed a measurable tail (1--2\% of the total) in the
distribution of position differences, with them being greater than 
0.5\,arcsec, a size not
seen with the SDSS data. Inspection of the images in this tail showed
many of these objects were not single stars, but galaxies or binary
stars. We therefore imposed a further check on the USNO-B guide stars,
that their positions had to be consistent with the corresponding 2MASS
data (after adding the systematic offset to put them on the USNO-B
system) to within 0.5 arc seconds. We did not apply any image
classification test to the USNO-B stars as the classifications were
less reliable than for SDSS and, at these magnitudes, the majority of
images are stellar.

The guide star magnitudes for 2dF must be in the range 12--14.5\,mag,
and must not differ in brightness by more than 0.5 mag to not exceed the dynamic
range of the guide camera. Since the WiggleZ project is conducted largely 
in the dark of moon, we opted for stars in the faintest range $14<r<14.5$ to
obtain as many as possible per field. In order to achieve good
astrometry it is also important to avoid stars with high proper
motion. We checked this using (USNO-B) catalogue data when available, but
as a further check, we also avoided very red stars, as many of these
are nearby M-dwarf stars which are more likely to have significant proper
motions.  The guide star selection criteria for the two survey regions
are summarised in Table~\ref{tab-guides}.

\subsubsection{Visual Inspection}

Finally, all the potential blank sky and guide star positions were
inspected visually by team members, using SuperCOSMOS \citep{Hambly2001}
images at each position. This step is necessary because our input
catalogues (like all automated image catalogues) suffer from a small
rate of object mis-classification and even missing objects. These
mistakes are not statistically significant, but could potentially ruin
a complete observation.

For the sky positions, the visual check is simply to ensure there
really is no object near the nominated position. The check is more
demanding for the guide stars: we must ensure that they are really
stellar, not multiple, and do not have any close neighbours (within
about 30 arc seconds).

A total of some 50,000 different sky and guide star positions were
defined across the different regions of the WiggleZ Survey, all of
which were visually checked by a team member. To speed this process
up, we designed a web-based interface that displayed large sets of SuperCOSMOS
images, 1000 or more at a time, and allowed the team member to easily
identify the small number of problematic positions.

\begin{table}
\caption{The selection criteria for guide stars}
\label{tab-guides}
\begin{tabular}{lll}
\hline
Criteria & SDSS  &  RCS2 (USNO-B)  \\
\hline
Magnitude & $14\le r \le 14.5$ & $13.95\le R \le 14.45$ \\
Colour    & $-0.5\le (g-r) \le 1.5$ & $-0.3\le (B-R) \le 1.7$ \\
Classification& stellar & any \\
Proper Motion& \multicolumn{2}{c}{$p<0.015$ arc seconds yr$^{-1}$} \\
\hline
\end{tabular}
\end{table}

\subsection{2dF  field placement}

When the WiggleZ survey began taking data with AAOmega, the GALEX data
set that formed the input catalogue for our survey was patchy and had
a highly varying density of targets.  Although both of these factors
have since improved tremendously, we do not use a simple grid of 2dF
field placements as it invariably produces highly pathological fields
that the 2dF robotic positioner is unable to configure in sufficient
time (i.e.\ less than the 1-hour typical observing time for each
field).

\begin{figure*}
\epsfig{file=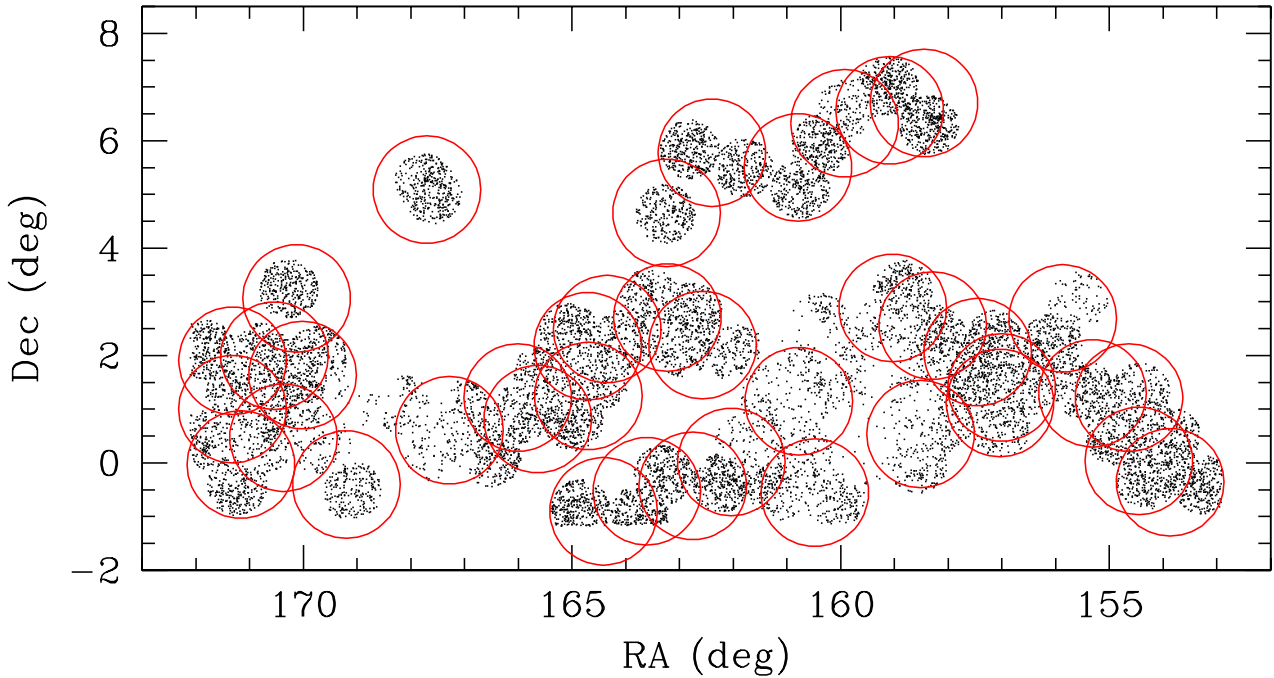,width=\figwidth\linewidth,angle=0}
\vspace*{-8.5cm} 
\caption{2dF field placement for the 15hr rectangle resulting from
  simulated annealing prepared for the 2008 April observing run.
  The target set is non-uniform with significant areas of GALEX
  targets yet to be observed (blank areas) and a number of targets
  within the available GALEX tiles having already been observed in
  previous runs (e.g.\ note the apparent under-dense area at
  RA$\sim$160 and Dec$\sim$1).  The simulated annealing approach
  ensures that a minimum number of 2dF positions is used to observe
  the targets.}
\label{fig-anneal}
\end{figure*}

Instead, we apply a simulated annealing algorithm to the problem of 2dF
field placement.  The algorithm used is identical to that
used by \citet{Campbell2004} \citep[see also][]{Metropolis1953} for
the 6dF Galaxy Survey \citep[e.g.][]{Jones2004}, but modified for use
with the 2dF system.  Briefly, the algorithm generates an arrangement
of 2dF placements that ensures that the smallest number of 2dF fields
are used to give as large as possible coverage of the target set with
as little variation in surface density as possible
(see Fig.~\ref{fig-anneal}).

The annealing is performed prior to each observing run in order to
have a fully efficient 2dF tile set to observe with.  In practice, one
issue that regularly arises with this process is that the annealing
algorithm simply aims to fill each 2dF field with $\sim$350\footnote{We enter
  350 fibres as the typical number of target fibres available per 2dF
  field.  Combined with 8 guide star fibres, up to 25 sky fibres for
  sky subtraction and a variable number of broken fibres at any one
  time, 350 target fibres represents a good estimate as to the
  available number of fibres.}  fibres.  In regions where the target
density is high such that there are many more targets than the 350  
the annealing algorithm is asked to assign fibres to, the result can 
potentially be a 2dF field where there
is a large target density gradient within that field.  This is
especially true of areas that consist of a single, non-contiguous
GALEX tile (Fig.~\ref{fig-anneal})---the annealing algorithm does not
know a priori to place the 2dF field such that it is centred on the
GALEX tile's centre in order to make the fibre positioning run faster.
Therefore we perform a small amount of manual intervention to optimize
the resultant 2dF field placement by, for example, putting single GALEX
fields in the middle of a 2dF circle.  Our later publications
\citep[e.g.][]{Blake2009b} will detail how we take account of the
overlapping regions for the window function of our survey.

\subsection{Observing priorities}

The WiggleZ survey has seven equatorial fields roughly
distributed over the full range of R.A. thus there are target fields
readily available throughout the year. Of course, this will change as
the survey progresses since more time will have to be expended on
observations of the more incompletely sampled target regions. The
target selection within a given WiggleZ survey region is necessarily
complex due to the rapid evolution of our data set and the complicated
selection criteria (Section~\ref{sec-selection}). However, choosing an
individual field pointing for a given night is mostly based on
minimizing its hour angle over the course of the observation. Other
factors typically considered when prioritizing field centres to be
observed, include: (1)\,the field is part of a larger contiguous area in a
WiggleZ region, (2)\,it contains a high target density (i.e., it will use all
of the fibres), and (3)\,it does not lie on the edge of a GALEX tile
or observed area (anticipating the GALEX data will be more contiguous 
in future observing runs).

\subsection{Final target lists and calibration}
\label{sec-final}

Once the field pointings have been selected for a night of observing,
combined target files for each field pointing are generated using
custom software (developed by RJJ). These files combine the target
galaxies (with suitable priorities assigned; see
Table~\ref{tab-priority}), guide stars, and blank sky positions. The
software also reads the lists of objects observed earlier in the run
to avoid repeat observations in overlapping regions. For the full
2-degree diameter field of view we generate a target file with an
optimal number of 700--800 possible targets.

Our target catalogues are updated before each observing run to record
which galaxies have already been observed. These are not normally
re-observed with the exception that targets that were previously
observed in poor conditions (typically with cloud cover more than 4/8
or seeing FWHM more than 3 arc seconds, resulting in a significantly
reduced redshift completeness, less than 50 per cent) without
obtaining an acceptable redshift are re-observed (see
Table~\ref{tab-priority}). Galaxies which were previously observed in
good conditions are put in a reserve list and are occasionally
re-observed at low priority ($P=2$ if no redshift was obtained; $P=1$
if a redshift was obtained).

A small number of additional calibration targets are included in the
target file at high priority. First, where possible we include two
standard stars: white dwarfs or hot sub dwarfs from the SDSS
compilation by \citet{Eisenstein2006}. These are used to assist in
future spectrophotometric calibration. Secondly, in order to determine
the reproducibility of our measurements, we include repeat
observations of three previously allocated targets and three
previously redshifted objects.

In regions where we are unable to use all the fibres on
high-priority WiggleZ targets, we allocate a small number of targets
for companion projects to the main survey: candidate cluster galaxies
from the RCS2 project and candidate radio galaxy identifications from
the FIRST \citep{Becker1995} survey. These additional targets are placed at lower
priority than all the new WiggleZ targets ($P=3$) and, in total, they
account for less than 1 per cent of the observations made.

Finally, we note that a small number of the spectra from each plate at
any time (up to 5 per cent) are affected by a fringing pattern
superposed on the target spectrum. We are unable to measure redshifts
for spectra strongly affected by the fringing.  The interference
fringing is caused by small air gaps between the ends of the 2dF
fibres and the 90 degree prisms that collect the target light on the
focal plane.  The air gaps create etalons, introducing interference
fringes which modulate the fibre transmission profile as a function of
wavelength.  The air gaps are the result of thermal expansion between
the steel fibre mounts and the glass fibres themselves.  The gap size
is mechanically unstable, relaxing during each exposure after being
disturbed during fibre placement, so the fringes are not removed by
the flat field correction in the data reduction process. A hardware
solution to the problem has been identified by the AAO, and will be
applied in due course.

Allowing for the fringing fibres, as well as those devoted to
background sky measurement and the various calibration measurements,
the average number of fibres devoted to new targets per observation in
the survey data is 325 (as quoted in Table~\ref{tab-design}.

% Database numbers at 2009 May 18; priority of the extras has been
% lowered, so these are upper limits for the final survey statistics.
% Ntotal    all,good = 376848,113811
% N_X RCS2  all,good = 3085, 2099 
% N_Y FIRST all,good = 1043, 97
% N_GALEX   all,good = 5, 4

\subsection{Field configuration}

The 2dF {\sc configure} software \citep{Shortridge2006} is then used
to generate configuration files from the target lists. This optimally
allocates all the fibres in use on the telescope to our targets. 
The configuration central wavelength is set to 670 nm due to our use
of the new WiggleZ dichroic beam splitter. Within the {\sc configure}
program, we generally choose to weight the peripheral fiducial
targets, use 25 sky fibres in a pointing and enforce a sky fibre
quota.  A simulated annealing algorithm incorporated into recent
versions (after v.7.3) of the {\sc configure} code
\citep{Miszalski2006} has greatly improved the yield of high priority
targets and the spatial uniformity.  Furthermore, high speed computers
installed in the AAT control room in 2007 October, allow us to
calculate optimal configurations for a typical field in less than half
an hour.

The output of the {\sc configure} software is a configuration file
specifying the position of every fibre, which is then used by the 2dF
positioner to place all the fibres on the one of the two field plates
that is to be used for the observation. The efficiency and reliability of the
positioner was improved significantly during the first year of our
observations, leading to positioner configuration times of less than
one hour for all the fibres.  These short configuration times provide
greater observing flexibility, particularly during good observing
conditions. During these clear periods of good seeing we can then
expose for less than our conventional one-hour exposure, and still
obtain the appropriate signal-to-noise ratio required for successful target
redshifting.

\begin{figure}
\epsfig{file=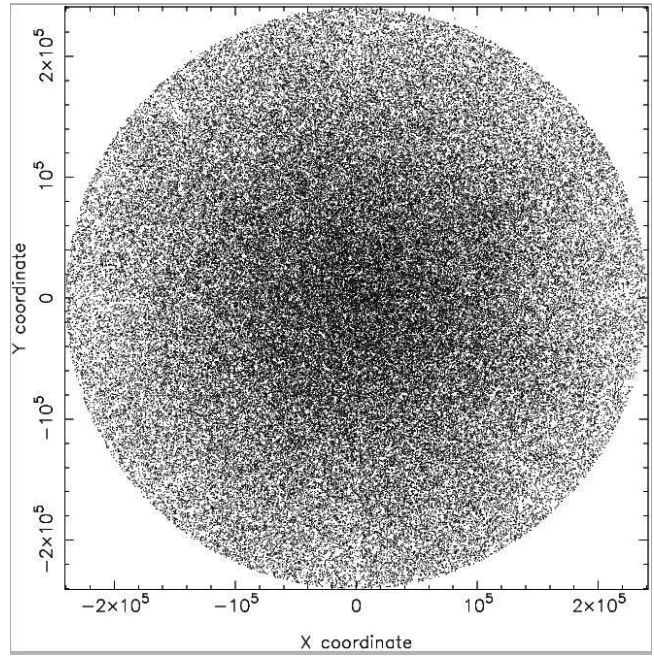,width=\figwidth\linewidth,angle=0}
\caption{The average distribution of fibres observed on the two 2dF field
  plates. Each dot represents one fibre observed in the first two
  years of the project excluding 2007 April--August when we only
  observed inside the central 0.7 degree radius. The smooth
  distribution (apart from very small regions at the edge due to the
  eight guide fibre bundles) demonstrates that no artificial spatial
  signal is introduced by the observing process.
}
\label{fig-fibres1}
\end{figure}

With a large number of observations now completed, we can illustrate
the success of this process by plotting the physical positions of the
fibres from all our observations. This is shown for both plates
combined in Figure~\ref{fig-fibres1}. Note that we have not included
data taken in the period 2007 April to 2007 August; during this period
we restricted our observations to a maximum radius of 0.7\,degrees from
the centre of each 2dF field as there was a problem with the
astrometric calibration of the 2dF positioner which reduced the
accuracy of fibre placement at larger radii.

The fibre distribution in Figure~\ref{fig-fibres1} reveals 8 areas of
slightly lower density around the edge of the field, associated with
the position of the 8 (equally spaced) fibre bundles used to measure
the fiducial stars. A similarly weak effect was present in the 2dF
Galaxy Redshift Survey (\cite[2dFGRS;][]{Colless2001} but only 4
fiducial stars in that case). The 2SLAQ galaxy survey \citep{Cannon2006}, 
by contrast, had a very marked discontinuity at the field edges
as only one of the 2dF spectrographs was used (alternate banks of
fibres were used for the companion quasar survey).  The fibre
distribution in Figure~\ref{fig-fibres1} also displays a strong radial
gradient in surface density. The average fibre density decreases by a
factor of about 3 from the centre to the edge of the field. This is a
natural consequence of the smaller size of the GALEX field compared to
the 2dF field: in many observations we do not fill the edges of the
2dF field with targets.

\subsection{Data reduction}

The AAOmega data are reduced during each observing run using the
automated {\sc 2dfdr} software developed at the AAO. We take standard
calibration flat field tungsten and arc (CuAr, FeAr, Ne and He) exposures 
for each pointing throughout the night, whose integration times are
typically 3\,sec and 30\,sec, respectively.  A two-dimensional bias frame is
constructed for the blue arm data to correct for a small number of
high bias columns on the blue CCD.

The data from the blue and red arms for a field pointing are reduced
separately, and then spliced together resulting in a final spectrum used
for target redshifting. The high-speed computers available at the AAT
make it possible to reduce the data in real time, allowing the
observers to have the majority of the data ready for redshift
measurement by the end of the night. We note that, as in the case of
most fibre-fed spectroscopy, we do not flux calibrate the
spectra. This is mainly due to the inherent difficulty in calibrating
the fraction of total light collected by each fibre from the
corresponding target. In a later paper, we will present the results of 
producing approximate flux-calibrated spectra from the data using
the white dwarf targets we include where possible in the SDSS fields
(see Sec.~\ref{sec-final}) and/or the optical photometry.

Significant refinement and further optimization of the {\sc 2dfdr} code 
has been undertaken by one of
us (SMC) and we have used these new versions of the software to have
the most reliable, efficient reductions possible for our observing
runs. Laplace filter cosmic ray rejection, Gaussian
spectral extraction, 1D scattered light filtering and optimal sky
subtraction have generally been used throughout the {\sc 2dfdr} 
reductions \citep[see][for
more details]{Croom2005}. Once the reduced, spliced file is generated
for a field, as a final reduction step we apply a principal
component analysis (PCA) sky subtraction algorithm to the spectra. 
This provides for better removal of the OH sky line contamination at the
red end of the spectra, thus making it easier to identify
redshifts. The PCA code we use was adapted (by KG and
EW) from software developed by \citet{Wild2005}. Our
standard approach for implementing the PCA sky subtraction is to
generate an eigenspectra file from the first 8--10 fields obtained on
an observing run. This eigenspectra file is then used to PCA sky
subtract the rest of the spectra obtained for a given run. We have
found the PCA sky subtraction is a useful step in that it
makes the redshifting a less onerous task by cleaning up the
``cosmetics'' of the spectra, particularly for data taken in good
observing conditions.

\section[]{Redshift analysis}
\label{sec-redshifts}

In this section we describe the process of redshift measurement from
our spectroscopic data, as well as various tests of the reliability of
these measurements.

\subsection{Measurement}

All redshifts in the WiggleZ survey are measured using an evolved
version of {\sc runz} --- the software used for both 2dFGRS
\citep{Colless2001} and 2SLAQ \citep{Cannon2006}.  The original
version of {\sc runz} made use of both discrete emission line redshift
determination and absorption line redshifts via template Fourier
cross-correlation \citep{Tonry1979}.  In the WiggleZ survey, it is
rarely possible to make use of absorption lines---the survey is
predicated on the ability to determine redshifts from emission lines
alone. Therefore, {\sc runz} has been modified significantly (by SMC),
optimizing it to measure redshifts from emission lines. The emission
line algorithm searches for sharp peaks in the spectrum, as candidate
emission lines, and then tries to match sets of these to known
combinations of strong lines at different redshifts.  The lines used
are: [OII]$\lambda$3727, H$\beta$, [OIII]$\lambda$4959,[OIII]$\lambda$5007, 
H$\alpha$, and [NII]$\lambda$6583. We note that the adopted wavelength for the 
[OII]$\lambda$3727 doublet is not the mid-point of the two transitions (which is
3727.42\AA), but rather is $\sim$0.4\,\AA\ redder to allow for the average ratio
found in our spectra.  For the cross-correlation measurement, {\sc
  runz} uses a revised set of template spectra including high-redshift
galaxies and QSOs, but the continuum level in typical spectra is so
low that the cross-correlation redshifts are very rarely used.

Uncertainties in the redshift measurements are estimated in two
different ways by {\sc runz}. In the more common case when the
redshifts are measured from emission lines, the redshifts are determined
for each individual line by fitting Gaussian profiles, using the
variance spectrum to determine the fitting errors.  The final emission
line redshift (and error) is the variance-weighted mean of that
derived from the individual lines.  For cross-correlation redshifts
the error is derived from the width of the cross-correlation peak
\citep{Tonry1979}.

% \begin{table}
% \caption{Template Spectra Used for Redshift Measurements}
% \label{tab-templates}
% \begin{tabular}{lll}
% \hline
% Name & Source & Description \\
% \hline
% NGC3379       &a & E1 galaxy \\
% NGC4889       &a & cD/E4 galaxy \\
% NGC5248       &a & SBbc galaxy (1) \\
% NGC2276       &a & SABc galaxy (1) \\
% NGC4485       &a & IBm galaxy (1) \\
% HD116608      &b & A1V star \\
% HD23524       &b & K0V star \\
% YALE1755      &b & M5V star  \\
% m3starSDSS    &c & M3 star (2) \\
% qso0          &d &  (3) \\
% qso05         &d &  (3) \\
% qso1          &d &  (3) \\
% qso2          &d &  (3) \\
% qso3          &d &  (3) \\
% newgdds-late  &e &  (1) \\
% newgdds-int   &e &  (1) \\
% newgdds-early &e &  (1) \\
% \hline
% \end{tabular}
% 
% Notes: (1) emission-lines in these galaxy templates were clipped prior to
% the cross-correlation analysis; (2) a template from the SDSS used to
% extend the M-stars to redder wavelengths; (3) the QSO templates are
% different parts of the same composite QSO spectrum split into more
% manageable portions to make the x-correlation faster. 
%  
% References to the sources of the spectra: 
% a=\citet{Kennicutt1992}
% b=\citet{Jacoby1984}
% c=\citet{Stoughton2002}
% d=\citet{Vandenberk2001}
% e=\citet{Abraham2004}
% \end{table}

\begin{figure}
\epsfig{file=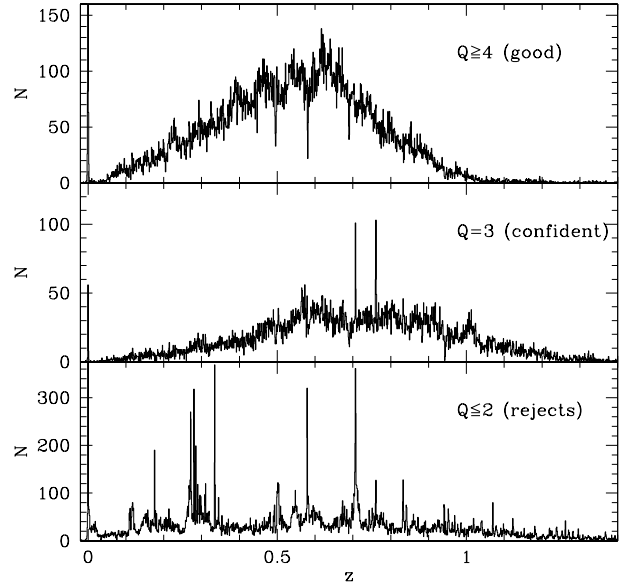,width=\figwidth\linewidth,angle=0}
\caption{Redshift distribution as a function of the quality flag
  $Q$. The top two panels are for accepted measurements ($Q$ of 3 and
  above). These show that objects at higher redshifts (with the
  H-$\beta$ line not visible) are generally not assigned the highest
  quality flags. The bottom panel is for rejected measurements ($Q\le
  2$): this has many sharp peaks in the distribution corresponding to
  misidentified sky lines. With two exceptions (see text), these sharp
  peaks are not evident in the good ($Q\ge 3$) redshift
  distributions.}
\label{fig-quality}
\end{figure}

In practice, complete automation of the redshift measurements has
proven to be problematic due to the noisy nature of many of the
spectra and the presence of artifacts such as residuals from imperfect
cosmic ray and sky removal and fibre fringing and cross-talk.  After a
redshift has been assigned using emission lines or the method of
\citet{Tonry1979}, {\sc runz} automatically generates an integer
quality flag ($Q$) in the range 1--5 based on how well the template
fits to the given spectra.  The user then inspects the given redshift.
In many cases, the user manually re-fits the spectrum and assigns a
new redshift and quality flag, in accordance with
the criteria in Table~\ref{tab-quality}.  In acknowledging that the
assignment of $Q$ values varies between users, we note that in any
future analysis only redshifts with values of $Q\ge 3$ will be
utilized.  Therefore, whilst there may be some debate as to what
constitutes the difference between $Q=4$ and $Q=5$, the critical
separation in quality occurs between $Q=2$ and $Q=3$.  This can be
seen in Figure~\ref{fig-quality} which displays redshift histograms
for both $Q\le 2$ and $Q=3$ redshifts for data collected up until 2008
March.  For the $Q=2$ redshift distribution, we see that there are
multiple peaks that coincide with assigning [OII] to sky lines---such
peaks are (with two exceptions) not present in the $Q=3$
distribution. Inspection of the two narrow peaks still visible in the
$Q=3$ data shows that the first ($z=0.707$) corresponds to
mis-matching [OII] to the weak sky line at 636.2 nm. The second peak
($z=0.761$) corresponds to mis-matching [OII] to rest-wavelength
Galactic H$\alpha$ emission at 656.3 nm which occurs in a few
fields. The number of spectra in these two peaks is less than 0.1 per
cent of the total sample.

\begin{table}
\caption{Redshift Quality Definitions}
\label{tab-quality}
\begin{tabular}{lrl}
\hline
Q & \% & Definition \\
\hline
1 & 21 & No redshift was possible; highly noisy spectra.\\
2 & 19 & An uncertain redshift was assigned.\\
3 & 18 & A reasonably confident redshift; if based on [OII]  \\
  &    & alone, then the doublet is resolved or partially resolved.\\
4 & 33 & A redshift that has multiple (obvious) emission  \\
  &    & lines all in agreement.\\
5 &  9 & An excellent redshift with high S/N that may be  \\
  &    & suitable as a template.\\
6 &  0.5 & Reserved for Galactic stars used as calibration sources.\\
\hline
\end{tabular}

Note: the second column lists the percentage of all observations
(including repeats) in each category.
\end{table}

\begin{figure*}
\epsfig{file=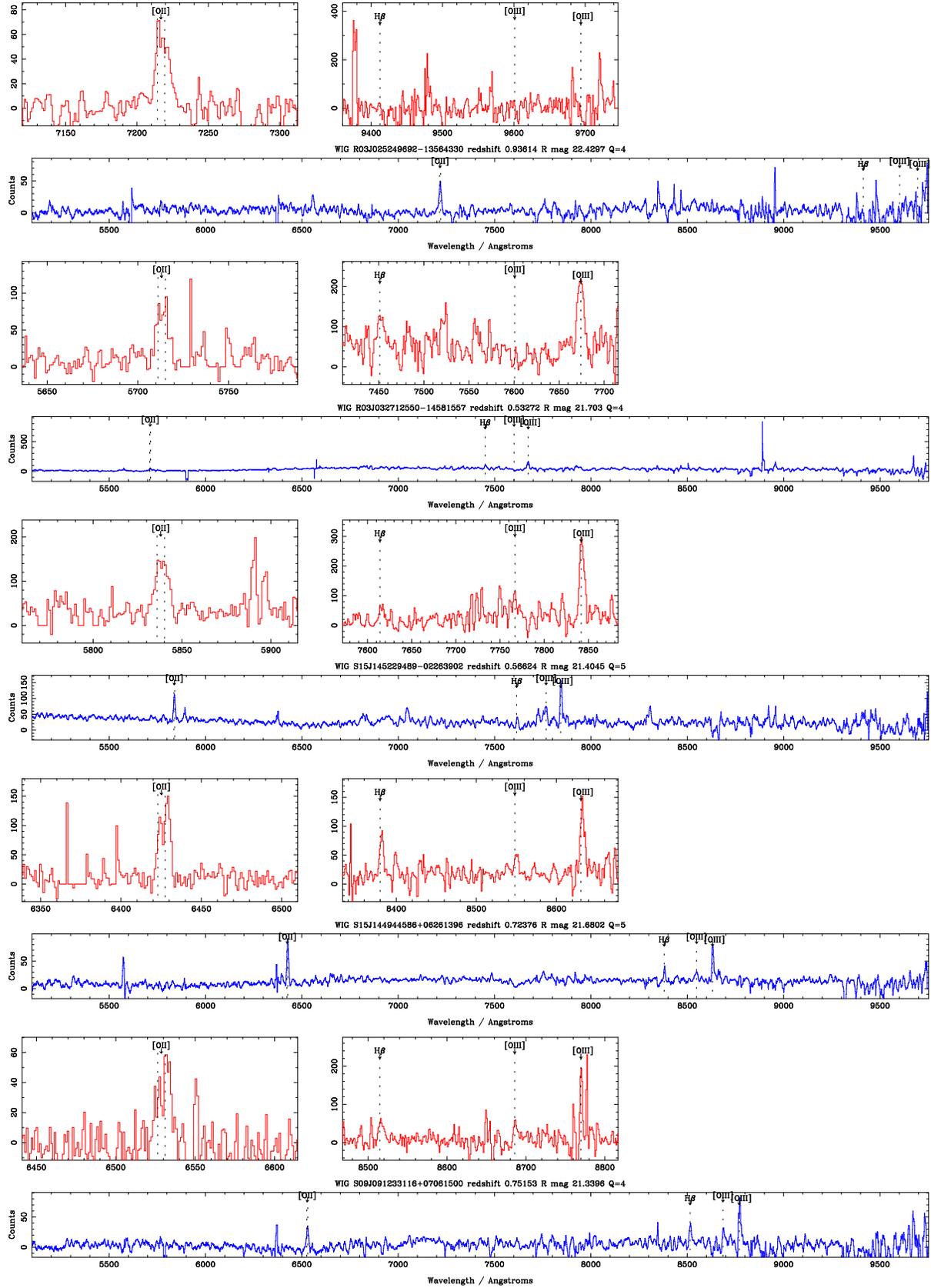,width=0.9\linewidth,angle=0}
\caption{Examples of randomly-selected high-quality (Q=4,5) spectra
  from the survey. For each spectrum, the upper panels (red) show the
  unsmoothed spectra zoomed in on the major emission lines, and the
  lower panel (blue) shows the whole spectrum, heavily smoothed in an
  optimal fashion. }
\label{fig-spectraQ45}
\end{figure*}

\begin{figure*}
\epsfig{file=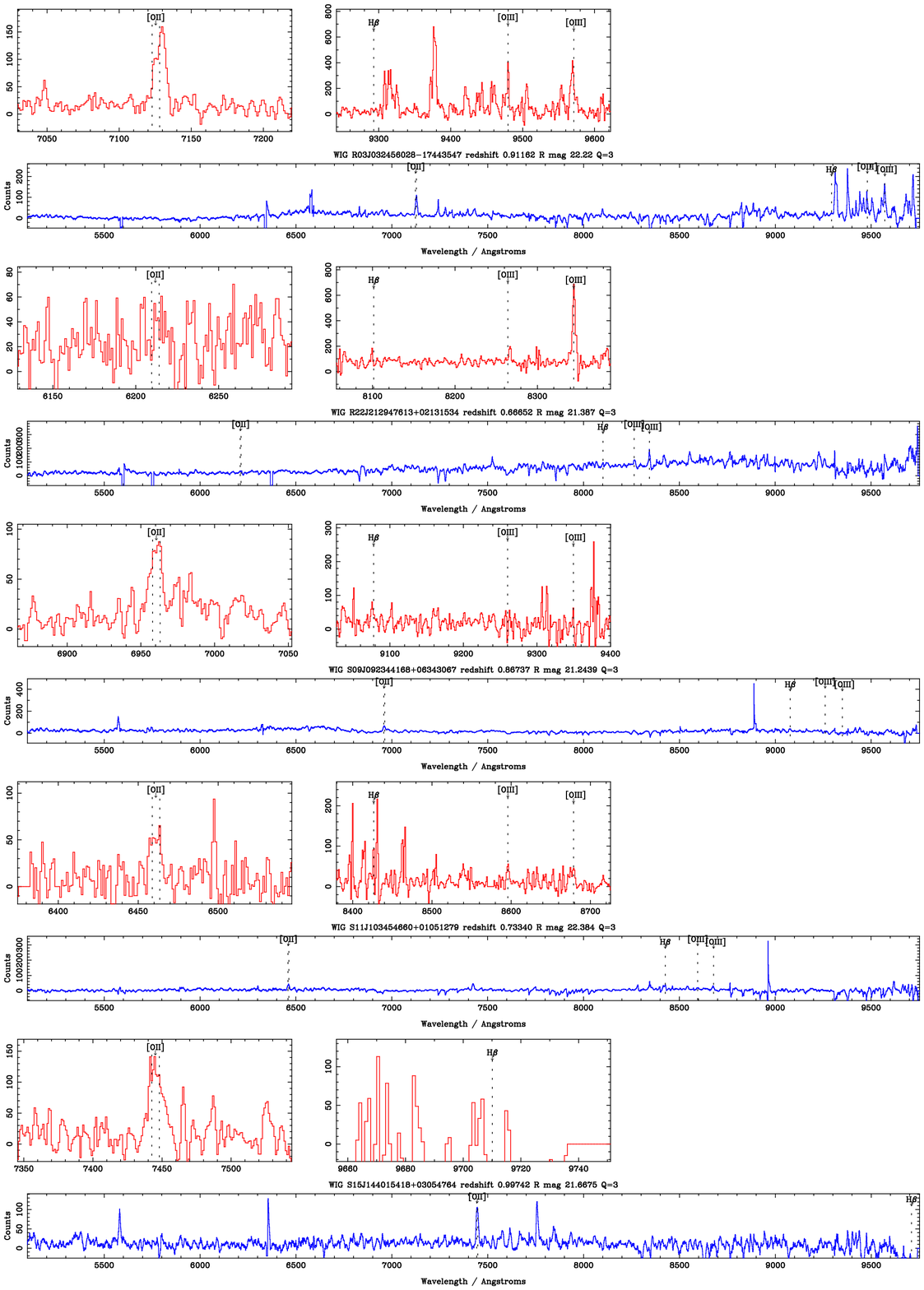,width=0.9\linewidth,angle=0}
\caption{Examples of randomly-selected low-quality (Q=3) spectra from
  the survey. For each spectrum, the upper panels (red) show the
  unsmoothed spectra zoomed in on the major emission lines, and the
  lower panel (blue) shows the whole spectrum, heavily smoothed in an
  optimal fashion.}
\label{fig-spectraQ3}
\end{figure*}

\begin{figure*}
\epsfig{file=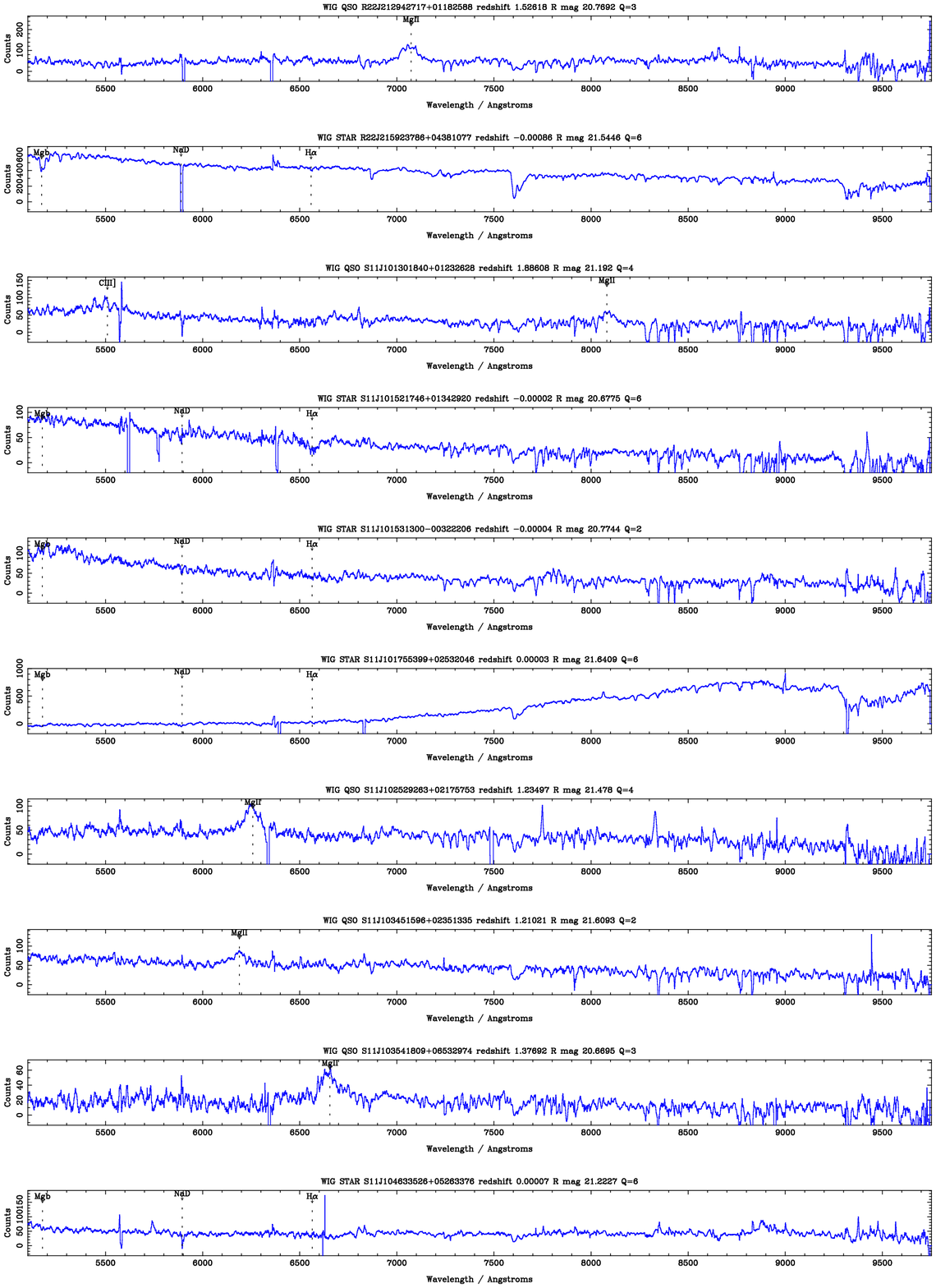,width=0.9\linewidth,angle=0}
\caption{Examples of randomly-selected QSO and star spectra from the
  survey. For each object the whole spectrum is shown, heavily
  smoothed in an optimal fashion.}
\label{fig-spectraXX}
\end{figure*}

We present examples of a variety of WiggleZ spectra in
Figure~\ref{fig-spectraQ45} (high quality Q=4-5), and
Figure~\ref{fig-spectraQ3} (low quality Q=3).  The spectra include
examples of those providing single [OII] doublet redshifts and standard 
multiple emission line detections. In Figure~\ref{fig-spectraXX} we
also present a random selection of QSO and star spectra from the
survey. This demonstrates the relatively low redshifts of the few QSOs
selected in the survey; many are only identified from the MgII
emission line.

\subsection{Redshift Reliability}

The one hour exposures that the WiggleZ survey uses are too short to allow
any significant detection of the continuum, with a continuum S/N
$\sim$ 1 per (approx 0.14 nm) pixel. However, by design, we are able
to reliably and accurately measure redshifts from the multiple bright
emission lines in the WiggleZ galaxies. We commonly detect the 
[OII]$\lambda$3727, H$\beta$, and [OIII]$\lambda\lambda$4959, 5007 
emission lines in the spectra. At
low redshifts (z $< 0.26$), when the [OII] line moves outside of our
wavelength range this is compensated by the ability to detect
H$\alpha$ in our spectra. At higher redshifts (z $>$ 0.95) we
often rely on redshifts obtained from a single line; however, the
spectral resolution is sufficient to see this line significantly
broadened at these redshifts (the [OII] doublet is often resolved at z
$\geq$ 0.8) which allows us to confidently conclude this single line
is [OII] and obtain the corresponding redshift. The reliability and
accuracy of using the broadening to confirm a single line in a
spectrum as the [OII] line of a high redshift emission line galaxy have
already been demonstrated by \citet{Breukelen2007}.

\subsubsection{Internal Tests}
\label{sec-internal}

We have multiple $Q\geq 3$ redshift measurements for over 6800 of the
galaxies observed to date, for both the SDSS and RCS2 regions. These
multiple ``correct'' redshift measurements can be used to estimate the
accuracy of our redshift measurements, and assess the effectiveness of
our redshift quality scheme. For the following analysis we define
redshifts to be reliable (i.e.\ the correct lines have been
identified) if they agree within $|\Delta z| < 0.002$. When converting
to velocity differences, we calculate these according to $\Delta v = c
\Delta z / (1 + z)$.

\begin{figure}
\epsfig{file=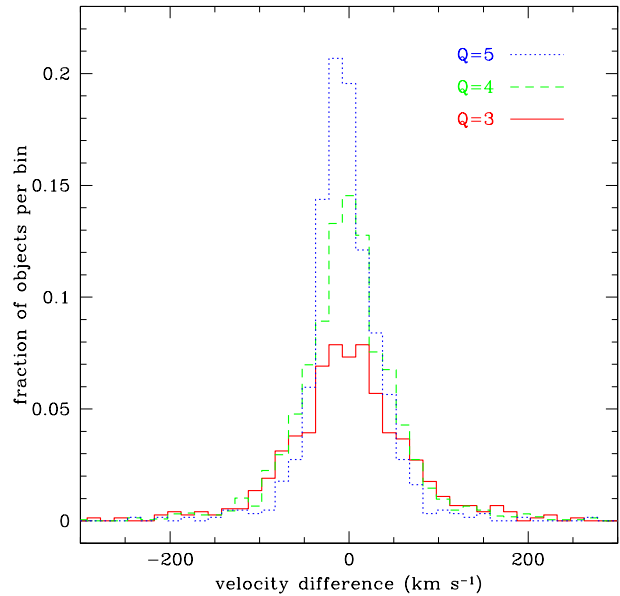,width=\figwidth\linewidth,angle=0}
\caption{Redshift differences from repeat observations of WiggleZ
  targets. The three histograms show the distribution of redshift
  differences in pairs of repeated WiggleZ observations. The solid,
  dashed and dotted curves are for pairs in which the qualities of the
  both measurements are 3, 4, and 5 respectively.}
\label{fig-zdiff}
\end{figure}

We start by assessing the internal agreement within each of the three
acceptable quality categories, $Q=3$--5, by taking objects whose
original and repeated measurements both have the same $Q$ value. There
are 737 $Q=3$ repeats, 2263 $Q=4$ repeats, and 619 $Q=5$ repeats. The
fractions of the redshift pairs that agree within each category are 68.5, 95.1,
and 100 percent respectively. For the correctly identified redshifts,
the mean difference is not significantly different from zero in any of
the categories, and the standard deviations are $\sigma_{\Delta z
  Q=3}=0.0042$ (68.5 \kms), $\sigma_{\Delta z Q=4}=0.0031$ (58.2
\kms), and $\sigma_{\Delta z Q=5}=0.0022$ (41.6 \kms)
respectively. (These values are for the differences of two
measurements so the uncertainty in an individual measurement would be
a factor of $\sqrt{2}$ times smaller.)  As expected, and as shown in
Fig.~\ref{fig-zdiff} the scatter in the redshift differences decrease
with increasing $Q$.

Given the high reliability of the $Q=4,5$ measurements, we can now
test a larger sample of $Q=3$ objects for which the repeated
measurement has a higher quality ($Q\ge4$) and can therefore be
regarded as correct. There are 2250 such objects, of which 78.7 per
cent are correct; the standard deviation in the redshift differences
is $\sigma_{z Q=3-4/5}=0.0039$ (67.5 \kms). The fraction of reliable
measurements is consistent with the internal value (68.5 per cent)
reported above. Similarly, 1425 $Q=4$ objects are repeated by higher
$Q=5$ measurements, of which 98.3 are correct and $\sigma_{z
  Q=4-5}=0.0029$ (54.0 \kms).  The average reliability and uncertainty
values for the three quality categories are summarised in
Table~\ref{tab-redshift}. These values are consistent with earlier
estimates made by \citet{Blake2009a} for a subset of the current data.

\begin{table}
\caption{Reliability and Uncertainty of Redshift Measurements}
\label{tab-redshift}
\begin{tabular}{crcc}
\hline
Quality  &  Reliability & \multicolumn{2}{c}{Uncertainty}  \\
    $Q$  &              & $\sigma_z$ & $\sigma_v$ (\kms)   \\
\hline
3  &  78.7 \%  &  0.00030  & 48.4 \\
4  &  98.3 \%  &  0.00022  & 41.2 \\
5  & 100.0 \%  &  0.00016  & 29.4 \\
\hline
\end{tabular}

Notes: the reliability fractions were obtained by comparing with
repeated measurements at higher quality ($Q$) values. The measurement
uncertainties $\sigma z, \sigma v$ were obtained by scaling the internal standard
deviations by $1/\sqrt{2}$. The velocity differences for each measurement were
calculated as $\Delta v = c \Delta z / (1 + z)$.
\end{table}

The comparisons of the repeat observations may also be used to check
the internal redshift uncertainties $\sigma z_{runz}$ for each
measurement estimated by the {\sc runz} code. Those internal
uncertainties have a median value of $\sigma z_{runz}\approx 0.00015$
for all three quality bins combined, but the distribution is very
non-uniform, with a long tail to higher values. Considering just this
median estimate of $\sigma_z =0.00015$, we find that it corresponds
well to the $\sigma_z = 0.00016$ derived for the $Q=5$ measurements in
Table~\ref{tab-redshift}. The empirical results for the $Q=3,4$
uncertainties in the table are larger than this, so in these cases the
internal redshift uncertainty may be an underestimate.

We also examined the redshift differences as a function of redshift,
for all galaxies with multiple $Q\geq 3$ redshift measurements. We do
not find any systematic variation with redshift of the mean or
standard deviation of the differences. We conclude that the redshift
uncertainties quoted above are valid over the entire redshift range of
the WiggleZ survey. This is, however, a trend in the reliability of
the $Q=3$ redshift measurements: the reliability decreases from about
80 per cent to about 40 per cent between redshifts of $z=0.8$ and
$z=1.2$ \citep[see Fig.~6 of][]{Blake2009a}.

\subsubsection{External Tests}
\label{sec-external}

At the time of writing, there are very few other galaxy surveys in our
redshift range that overlap with the WiggleZ survey, so there are few
opportunities for external comparison. There is a small overlap
between WiggleZ and Data Release 3 of the DEEP2 survey
\citep{Davis2003} in our 22-hour field. Although the DEEP2 survey (in
that region) is aimed at $z>0.5$ targets, there are 53 sources in
common with WiggleZ (positions matching to within 2.5 arc
seconds). The estimated r.m.s.\ uncertainty in the DEEP2 redshifts is
30 \kms with less than 5 per cent incorrectly identified
\citep{Davis2003}.

% For comparison data see /Users/mjd/Documents/Research/wigglez/analysis/redshifts

\begin{figure}
\epsfig{file=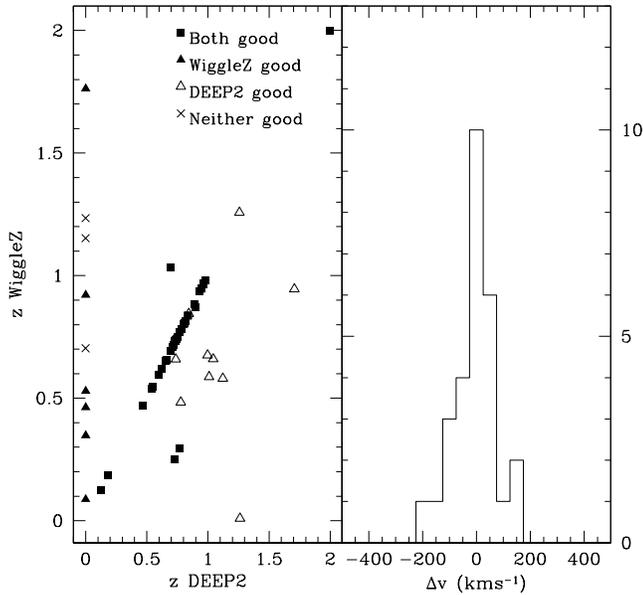,width=\figwidth\linewidth,angle=0}
\caption{Comparison of WiggleZ redshifts with DEEP2 survey data. The
  left-hand panel compares the different redshifts with the symbols
  indicating the reliability status of the measurements in the
  respective surveys. The right-hand panel shows the histogram of
  redshift differences.}
\label{fig-deep2}
\end{figure}

We compare the WiggleZ and DEEP2 redshifts in Fig.~\ref{fig-deep2}. Of
the 34 sources that both surveys assign good redshifts to, only 3
disagree substantially (due to different line identifications).  Of
the redshifts that disagree, all three are $Q=3$ WiggleZ measurements;
no $Q=4$ or 5 WiggleZ measurements disagree with the DEEP2
measurements. On careful inspection of the WiggleZ spectra and the
corresponding DEEP2 spectra we conclude that in two cases the DEEP2
value is correct; the third case is ambiguous so we cannot determine
which is correct. This rate of misidentification among the $Q=3$
WiggleZ redshifts (20-30\% of the $Q=3$ redshifts are wrong) is
entirely consistent with the internal estimates in
Section~\ref{sec-internal} above.

Restricting the sample to those that agree within $|\Delta z|<0.002$,
there is no significant mean difference ($\overline{\Delta
  z}=-0.00003\pm 0.00002$) and the rms difference is $\sigma_{\Delta
  z}=0.00047$ (80.5 \kms). This scatter is entirely consistent with
the estimates of the internal uncertainties in the respective surveys
quoted above.

\section{Initial  Results}
\label{sec-results}

The primary aim of our project --- measurement of the baryon acoustic
oscillation scale --- cannot be completed without the full data
set. However there are many secondary scientific projects possible
with the data currently available. In this section we describe some
initial results from the current data sample. We focus on the
properties of the measured galaxies, noting in particular the success
of our methodology in selecting high-redshift galaxies.

\subsection{Success of Target Selection}

The whole WiggleZ survey strategy is based on the ability to select
high-redshift emission-line galaxies for the spectroscopic
observations. We initially selected galaxies based on the UV and
UV-optical colours, but then improved the selection with optical
colour selection to remove some of the low-redshift targets as
described in Section~\ref{sec-selection}. The success of this strategy
is demonstrated in Figure~\ref{fig-selection} by an increased fraction
of high-redshift targets in both the SDSS and RCS2 regions. In the
SDSS fields the median redshift increases from $z_{med}=0.53$ to
$z_{med}=0.61$. In the RCS2 fields, the more accurate optical colours
permit a greater improvement: the median redshift increases from
$z_{med}=0.54$ to $z_{med}=0.67$. In the combined sample the median
redshift is $z_{med}=0.63$ and the redshift range containing 90 per
cent of the galaxies is $0.22 < z < 1.02$.

The ability of the survey to obtain the required number of galaxies is
also a function of the spectroscopic {\em success rate}, defined as
the fraction of {observations} for which we obtain a reliable ($Q\ge
3$) redshift. By comparison, the overall survey {\em completeness} is
defined as the fraction of all the input targets for which we obtain a
reliable ($Q\ge 3$) redshift.

{The mean spectroscopic success rate for the survey so far is 60 per
cent (Table~\ref{tab-design}); the completeness (after the repeated
observations) is 70 per cent. Longer exposures would enable a higher
success rate and better quality spectra but would result in greatly
diminished values for the total area and total number of galaxies
covered and hence a poorer BAO measurement.}

\begin{figure}
\epsfig{file=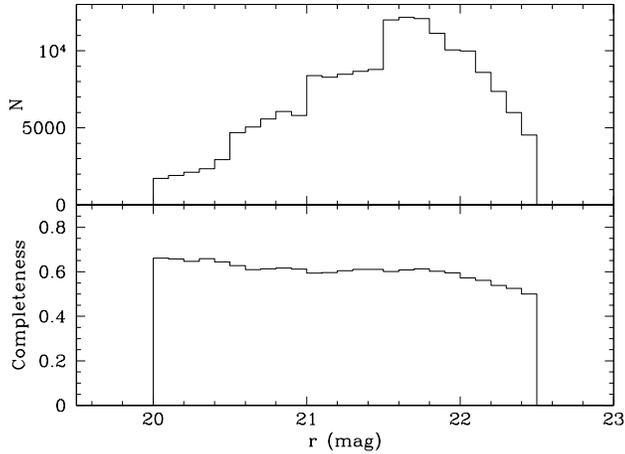,width=\figwidth\linewidth,angle=0}
\caption{Spectroscopic success rate of the WiggleZ survey as a
  function of $r$ magnitude. Top: the $r$-magnitude distribution of
  the observed targets. The steps are caused by our prioritisation
  scheme (see Table~\ref{tab-priority}). Bottom: success rate defined
  as the fraction of observed targets for which good redshifts were
  measured, shown as a function of $r$ magnitude.}
\label{fig-completeness}
\end{figure}

The main factor that strongly affects the success rate is the weather:
it drops by around 20 per cent when observations are
taken through cloud. (In the first half of the survey, 57 per cent of
the allocated time was clear and we observed through cloud for an
additional 9 per cent of the time.) In Figure~\ref{fig-completeness} we
show that the success rate is not a strong function of the continuum
$r$ magnitude. This is expected as we are measuring the redshifts from
emission lines and not from continuum features. The upper panel of
Fig.~\ref{fig-completeness} shows the distribution of $r$ magnitudes
for the targets: this shows steps at half-magnitude intervals due to
our prioritization scheme (see Table~\ref{tab-priority}).

\subsection{Galaxy Properties}
\label{sec-properties}

We will examine the physical properties of the WiggleZ galaxies in
detail in later papers, but some preliminary observations can be made
at this stage. 

The WiggleZ galaxy sample clearly favours blue galaxies with strong
emission lines as a result of the primary NUV flux limit and the
$NUV-r$ colour selection. This selection has been quantified by our
analysis of DEEP2 \citep{Davis2003} galaxies for which deep optical
and GALEX observations were available.  (see
Section~\ref{sec-selection}). We find that, as expected, the WiggleZ
targets are blue compared to optically-selected galaxies: 82 (99) per
cent of the WiggleZ targets are bluer than $B-R=1.2$ (1.8).

\begin{figure*}
\epsfig{file=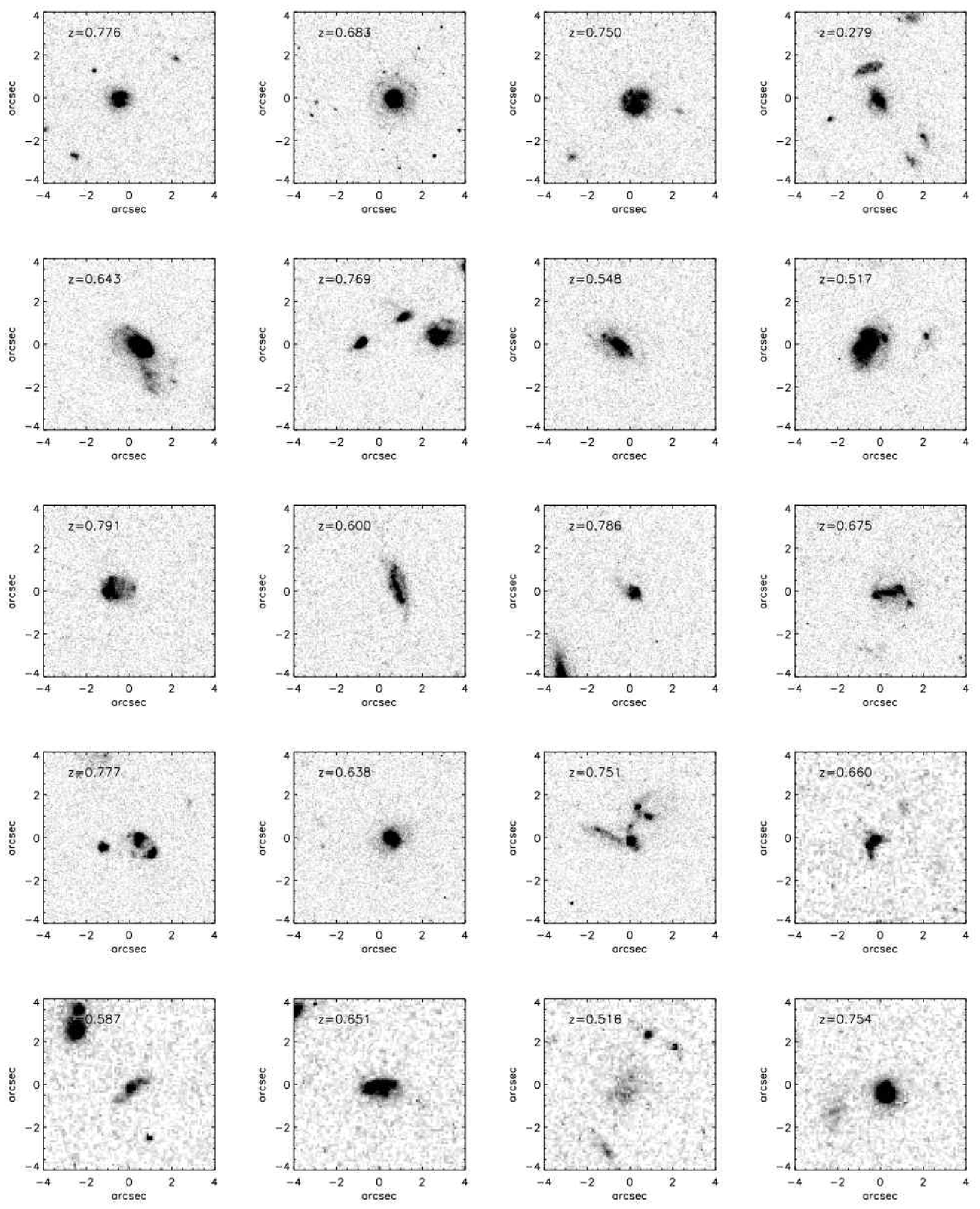,width=\figwidth\linewidth,angle=0}
\caption{High-resolution {\em Hubble Space Telescope} archive images of a
  selection of WiggleZ galaxies. Each image is 8 arc seconds on a
  side. These images demonstrate that a high fraction of the WiggleZ
  galaxies are interacting.
}
\label{fig-morphology}
\end{figure*}

We have investigated the physical morphology of the WiggleZ galaxies
by searching the {\em Hubble Space Telescope} archive for any images
that are available of the WiggleZ galaxies we have observed. We
present a selection of these images in Figure~\ref{fig-morphology}. The
images show that a very large fraction of the WiggleZ galaxies are
interacting or show other signs of a disturbed morphology. This
suggests that merger activity may play an important role in driving the
strong star formation detected in the WiggleZ galaxies. In terms of
the galaxy photometry, we also note that about 25 per cent of the
images reveal more than one galaxy within the $\approx 4$ arc second
PSF of GALEX. In these cases the UV photometry may include both sources
whereas the optical measurements may include only one of the sources.

\begin{figure}
\epsfig{file=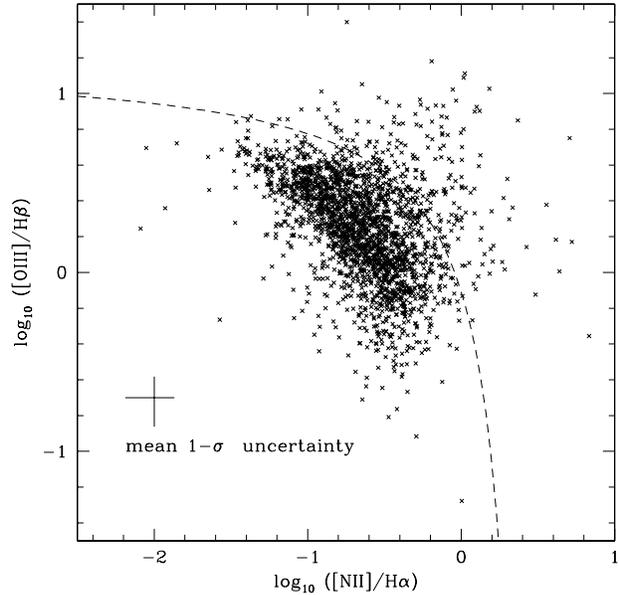,width=\figwidth\linewidth,angle=0}
\caption{Emission line ratios of WiggleZ galaxies. In this
  ``diagnostic'' plot the dashed line indicates the Kewley et al.\
  (2001) extreme star formation envelope. Any galaxies above the
  dashed line are dominated by AGN activity and not star
  formation. The diagram shows that the WiggleZ galaxies are
  predominantly star-forming galaxies. The plot is based on a
  low-redshift subset of the WiggleZ sample because we cannot measure
  [NII] and H-$\alpha$ emission at redshifts above $z\approx 0.4$.  }
\label{fig-bpt}
\end{figure}

The most direct physical measurement we have for each WiggleZ galaxy
is from the various emission lines in the spectra, notably [OII] which
is the one line present in nearly all spectra. At low redshifts
($z\lesssim 0.35$ for the old dichroic and $z\lesssim 0.48$ for the
new dichroic), both H$\beta$ and H$\alpha$ are also visible. This is
below the median redshift range for the survey, but it does provide a
sub-sample of galaxies for which we can investigate the line
diagnostics. In Fig.~\ref{fig-bpt} we show the ``BPT''
\citep{Baldwin1981} line ratio diagnostic plot for low-redshift
galaxies in our sample. Note that the two lines in each ratio
([OIII]/H$\beta$ and [NII]/H$\alpha$) are at very similar
wavelengths, so the plot can be constructed from spectra that have not
been flux calibrated.  We also show in the figure the ``maximal star
formation'' boundary line \citep{Kewley2001} between the star formation
(below the line) and the active galaxy (above the line) domains. The plot
shows most (90 per cent) of the WiggleZ galaxies fall below the
nominal boundary line, confirming that we are mainly selecting
star-forming galaxies as intended.

\begin{figure}
\epsfig{file=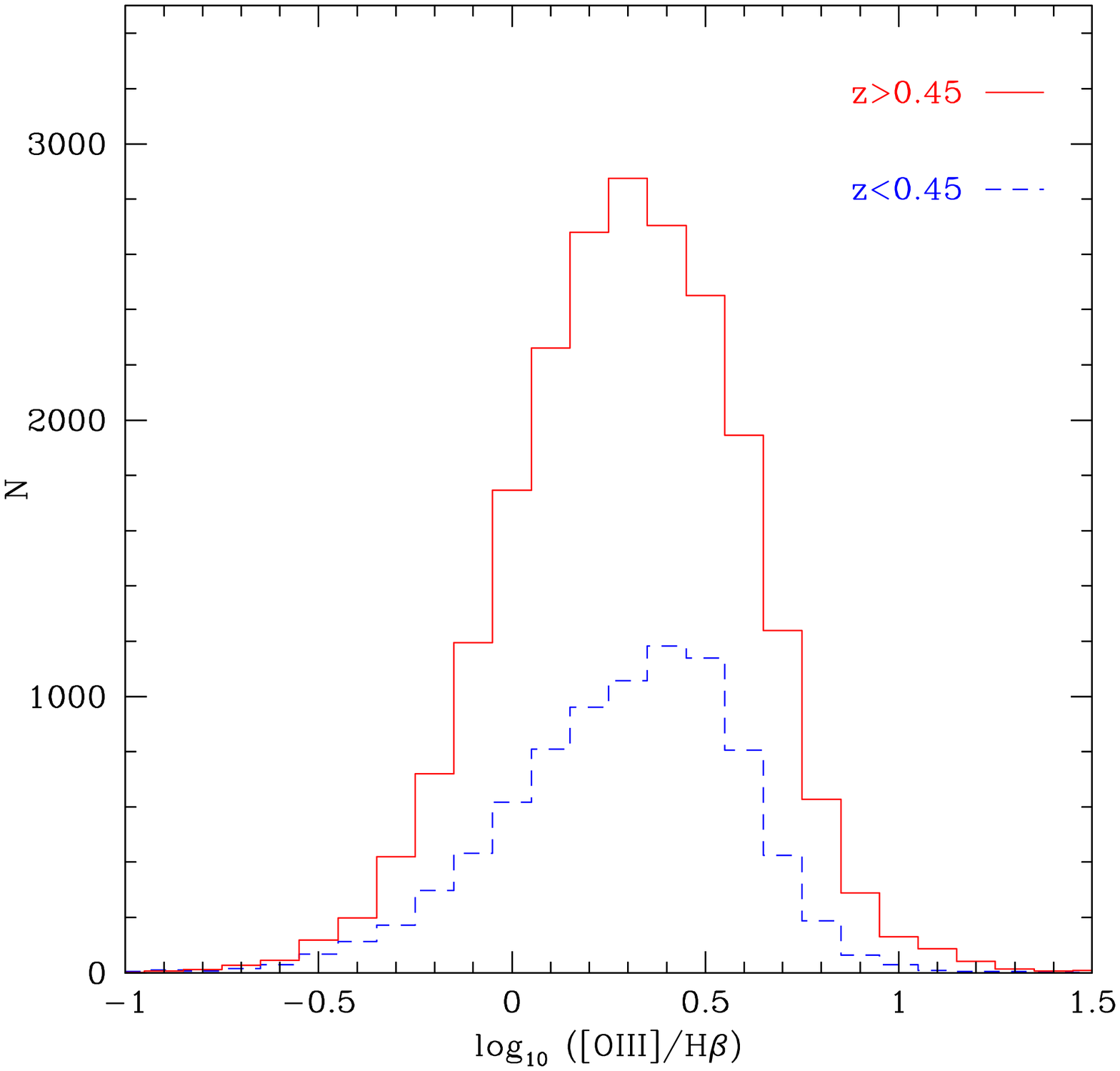,width=\figwidth\linewidth,angle=0}
\caption{A comparison of the [OIII]/H-$\beta$ line ratios of high
  ($z>0.45$) and low ($z<0.45$) redshift WiggleZ galaxies. Analysis of
  the low-redshift sample (Fig.~\ref{fig-bpt}) shows that it does not
  contain large numbers of AGN. This figure shows that the high
  redshift sample does not contain any large excess of AGN, as would
  be indicated by line ratios of log$_{10}$[OIII]/H-$\beta > 0.9$.
}
\label{fig-bpt2}
\end{figure}

We can also use the [OIII]/H$\beta$ line ratio to show that the
higher redshift targets are not dominated by AGN. In
Figure~\ref{fig-bpt2} we compare the distributions of the
[OIII]/H$\beta$ line ratio for galaxies above and below a redshift of
$z=0.45$. It is clear from the figure that the high redshift galaxies
do not have any large excess with high line ratios,
indicative of AGN. In the low redshift sample that we
analysed in Figure~\ref{fig-bpt}, the fraction with
log$_{10}$[OIII]/H$\beta > 0.9$ is 1 per cent, compared to 2 per cent
for the galaxies at higher redshifts ($z>0.45$).

\subsection{Galaxy Distributions}

\begin{figure*}
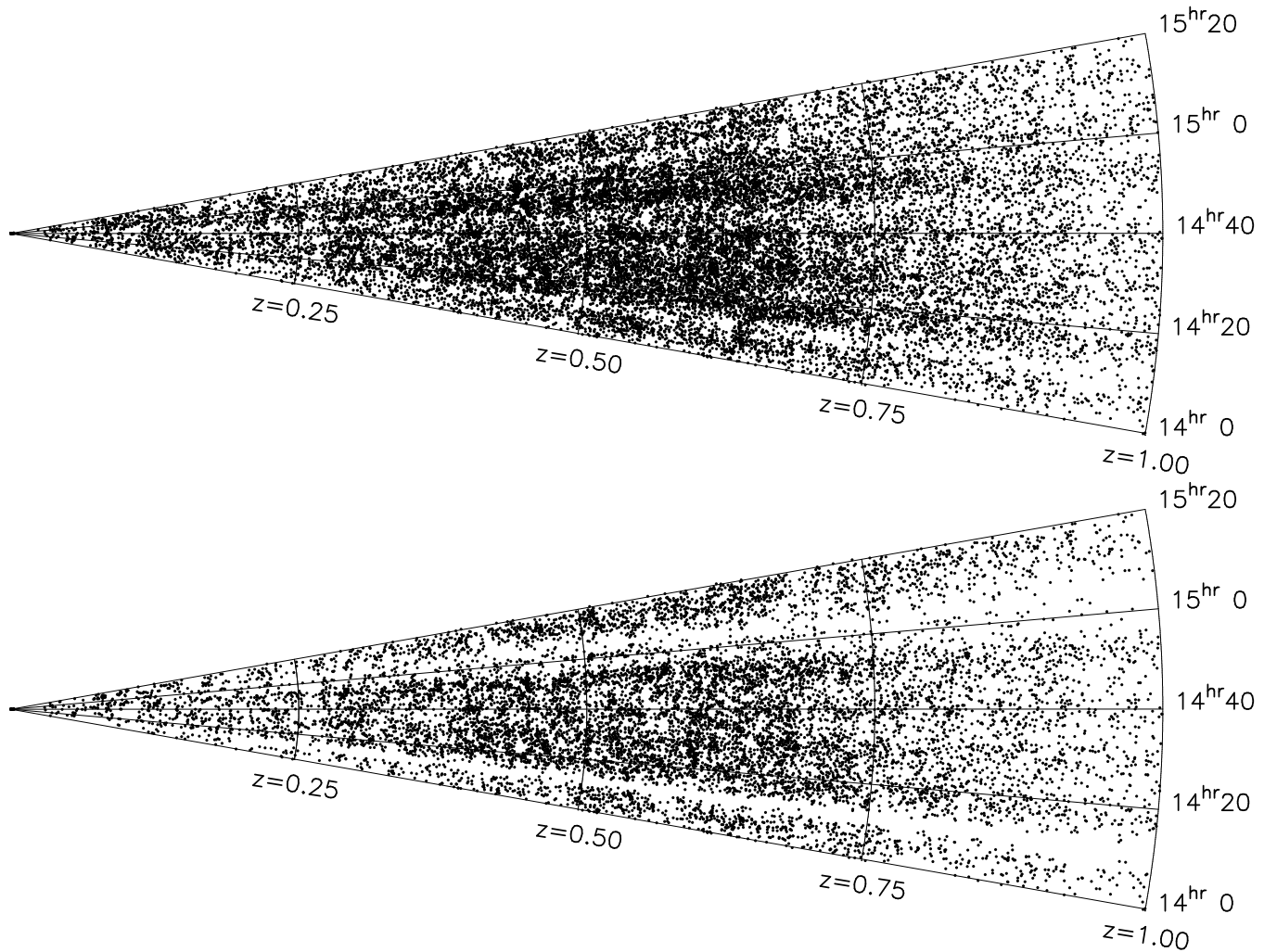

\epsfig{file=fig_cone.ps,width=\figwidth\linewidth,angle=0}
\epsfig{file=fig_cone_thin.ps,width=\figwidth\linewidth,angle=0}
\caption{The spatial distribution of WiggleZ galaxies observed in the
  15-hour region. Each plot shows the comoving positions of the galaxies 
  projected onto Right Ascension. The full width shown --- the
  comoving distance to a redshift of $z=1$ --- is 3.3\,Gpc. The upper
  plot shows all galaxies in the field. The lower plot only includes
  galaxies with Declinations greater than $+1$ degree (about half the
  sample) to make the small scale clustering easier to discern.
}
\label{fig-cone}
\end{figure*}

We show an example of the spatial distribution of the WiggleZ galaxies
from the 15-hour field as cone plots projected onto the Right Ascension plane in
Figure~\ref{fig-cone}. The large-scale radial features in the plots
are due to the incomplete coverage of the survey at present. In the
remaining regions of the plots, especially the lower plot, some
evidence of spatial clustering is revealed by visual inspection. This
is in contrast to similar plots for the 2dF Galaxy Redshift Survey
\citep{Colless2001} and the 2SLAQ Luminous Redshift Galaxy Survey
\citep{Cannon2006}, both of which show strong evidence of clustering
in similar diagrams. This difference is expected because the WiggleZ
galaxies are comparatively rarer objects with much larger median
separations than the other samples. Furthermore, we are surveying much
larger spatial scales. The distribution of WiggleZ galaxies probes
clustering on 100\,Mpc scales where it is weaker than on the much
smaller scales probed by, for example, the 2dFGRS.

\subsubsection{Small-scale galaxy clustering}

Although the WiggleZ galaxies are not as strongly clustered as red
galaxies, their intrinsic small-scale clustering strength is a crucial
design parameter for the main WiggleZ survey. We have used the early
survey data to measure this small scale clustering. We find that the
clustering is actually much stronger than that measured previously for 
UV-selected galaxies at low redshifts ($z\approx 0$). We measure a
clustering length of $r_0=4.40 \pm 0.12 h^{-1}$ Mpc
\citep{Blake2009a}. This is comparable to the Lyman break galaxies
measured in high redshift ($z\approx 3$) surveys; the combined flux
limits and colour selection of the WiggleZ survey are selecting
extremely luminous blue galaxies that are relatively strongly
clustered. This is highly advantageous for our survey design, as
discussed above.

We have estimated the final precision of the survey in measuring the
dark energy equation of state parameter $w$ assuming a model with
constant $w$ \citep{Blake2009a}. We include the 5-year WMAP cosmic
microwave background data \citep{Komatsu2009} and the latest supernova
data from the SNLS, {\em HST} and ESSENCE projects
\citep{Astier2006,Riess2007,Wood2007} to predict marginalised errors
in our model of $\sigma(\Omega_m)=0.02$ and
$\sigma(w_{const})=0.07$. Not only is this more precise than previous
estimates, but the method is independent and eliminates existing
degeneracy in the ($\Omega_m, w_{const}$) plane \citep[see Fig.~20
of][]{Blake2009a}.

\subsubsection{The Galaxy Power Spectrum}

We have also measured the three-dimensional power spectrum of the
WiggleZ galaxies on large scales. We find that a model power spectrum
based on linear theory is a good fit to the data over a wide range of
scales up to a maximum wave number of $k<0.25 h \Mpc^{-1}$
\citep{Blake2009b}. The fit extends to such large wave numbers because
of the high redshifts of the WiggleZ galaxies and their relatively low
bias. We also detect the imprint of peculiar velocities when comparing
the power spectrum measurements as a function of radial and tangential
wave numbers: the results are consistent with a redshift space
distortion parameter of $\beta=0.5$.

\section{First Public Data Release}
\label{sec-data}

This paper accompanies the first public data release from the WiggleZ
survey, made at the half-way point in terms of allocated observing
nights. In this section we summarise the current status of the
survey and give details of the survey database design and the
first public data release.

\subsection{Survey Progress}

At the end of the 2008A observing semester, the WiggleZ Survey reached
its half-way point in terms of scheduled nights on the telescope (see
Table~\ref{tab-dates}): 112 of a nominal 220 nights had been
scheduled. Several improvements were made to the survey design
parameters in the first twelve months as a result of the initial
observations, but the design and performance are now stable. 

The key parameters of the survey performance are summarised in
Table~\ref{tab-design}. The first part of the table lists the various
parameters we obtain for the survey which, when multiplied
together give us the rate of galaxies successfully
observed per clear night. In the second half of the table we calculate
the length of the survey needed in order to observe the 240,000
galaxies we need for the clustering measurements. This
comes to about 160 clear nights, corresponding to 220 nights when we
include the standard factor of 1.33 to allow for bad weather. 

\begin{figure}
\epsfig{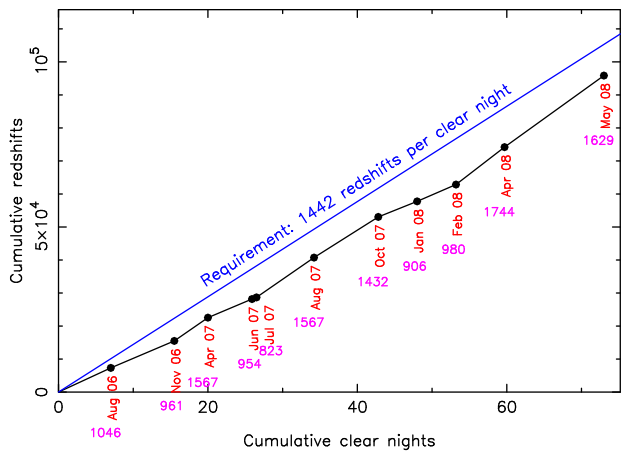}
\caption{Overall progress of the WiggleZ survey measured by the number
  of galaxies with redshifts measured per clear night. The
  target number of 1442 galaxies per clear night is based on our
  overall target of 240,000 galaxies given the standard clear weather
  fraction of 0.75 and a survey duration of 220 nights. Data points
  are shown for each observing run, with the number of redshifts
  measured per clear night in each run printed below each point.}
\label{fig-progress}
\end{figure}

As noted above, the total number of survey galaxies required to be
observed is a function of
the target selection and the observational success rate. Other factors
such as the number of fibres allocated per field and the number of
fields observed per night also enter into the calculation which is
summarised in Table~\ref{tab-design}. In Fig.~\ref{fig-progress} we
present the actual results of the observations graphically, showing
the cumulative number of survey galaxies successfully observed as a
function of the number of clear nights used. This is compared to the
required rate of 1442 survey galaxies per clear night to reach 240,000
galaxies in the full survey. As can be seen from the figure, the
survey is almost on schedule to achieve this goal.

\subsection{Database design and web application}
\label{sec-database}

The WiggleZ Survey is a complex, multi-year campaign involving the
collection of many nights of observational data. Thus a sophisticated,
reliable data archiving system is an absolute necessity. Our general
approach is to take the raw and reduced data, observational logs,
catalogue and other associated files and organize them within a well
defined directory structure on a large cluster server. An extensive
suite of software has been developed (by NJ) to serve the data from
this archive via a MySQL database and ``Ruby on Rails'' web
application. Python scripts have also been built to assist in the data
uploading, checking and management of the archive.

The web application gives access to data subsets using a simple search
interface which allows queries based on the observation dates, target
identifiers, regions in the sky, filter bandpasses, redshift range and
other parameters, or some combination thereof.  All these parameters
for the large number of targets have been captured in the
database. Users can also execute more complex, custom SQL queries on
the database via the web application. Catalogues of targets, and the
associated FITS files of the reduced spectra, can be downloaded for
analysis. Finally, users of the web application can also upload
catalogues of objects to be cross-matched against targets currently
residing in the WiggleZ data archive.

The WiggleZ web application and database have been designed to capture
all the pertinent data regarding the observations so that we can
analyze our large data set with as many approaches as possible, thus
maximizing the scientific return. The web application is very
user-friendly and well documented, such that a novice user can soon
execute complex queries of the database with minimal training. A
cookbook of possible web application uses is provided along with
visual explanations of the various features. Since the web application
has the flexibility of allowing both simple searches and more complex,
custom SQL queries, users can mine the database for the specific
results that they seek. After a successful query of the database the
user can then use the search results page to: select and view an
individual spectrum, select (sub-)sets of resultant targets for
download as a tar file of FITS spectra files or as a catalog file (in
ASCII, VOTable or FITS format). When viewing an individual spectrum
from a search, the associated GALEX and optical images of the target
(when either are available) are displayed, and the user has the
capability of zooming in on the spectrum, applying Gaussian smoothing
and downloading hard copy files (in FITS or postscript format).

The web application also allows a user to upload a space-separated
ASCII file containing RA and DEC positions of targets, to cross-match
this list with targets in the WiggleZ database. The results of the
cross-matching can be downloaded in the form of either ASCII, VOTable
or FITS table catalogues, or as a tar file of FITS files containing
the spectra. Finally, download requests of selected spectra or
catalogues from the search results page results in the user being
redirected to the ``pending requests'' page. On this web page,
multiple queries by a user can be left pending and downloaded at a
later time.

\subsection{Public Database Release}

The first public release of WiggleZ data (DR1) aims to provide the
astronomical community with an opportunity to utilize our large data
archive well before the survey is complete and before the final, complete
data release. This first data release is available via the following
URL: http://wigglez.swin.edu.au/. This web page does not require
secure access and is available to any user who wants to access
the DR1 data. 

The WiggleZ DR1 database provides the target coordinates, photometry
and redshifts, along with the quality of the determined redshifts,
downloadable in ASCII, VOTable and FITS table format. It should be
emphasized that the DR1 targets will also have their FITS spectra
files publicly accessible via the website.  We have included the UV
and optical magnitudes for all of the targets in DR1, where
available. The DR1 public data release includes all observations up to
the end of Semester 2008A: a total of 169,000 spectra with 100,138
successful redshift measurements. The median redshift is $z=0.6$ and
the redshift range containing 90 per cent of the galaxies is
$0.2<z<1.0$.

\section{Summary}
\label{sec-summary}

The WiggleZ Dark Energy Survey will be the first large-volume
spectroscopic galaxy survey to measure the BAO scale in the key
redshift range of $0.2<z< 1.0$. 
% {\it WJC: we seem to be inconsistent in the range we quote here, with
%  this range sometime used as well as $0.3<z<0.9$!}
In this paper we have described the
design and initial results of the WiggleZ Survey at the half-way point
which coincides with our first public data release.

To emphasise the large size of the WiggleZ survey we note that, as of
2008 December, we have measured 119,000 galaxy redshifts, of which 73
000 have redshifts $z>0.5$. This means we have already more than
doubled the number of known galaxy redshifts above $z=0.5$ from the
two largest existing surveys: DEEP2 \citep{Davis2003} with
$N_{z>0.5}\approx 28,000$ and the VIMOS VLT Deep Survey
\citep[VVDS,][]{Garilli2008} with $N_{z>0.5}\approx 18,000$. At the
time of writing, the total number of $0.5<z<1.5$ redshifts published
from all surveys is 100,000 (as listed by NED). The WiggleZ survey,
when completed, will more than double this quantity.

Our survey design relies on the use of GALEX UV satellite data to
efficiently select high-redshift emission line
galaxies. These objects, whilst fainter than giant red galaxies at the
same redshift, can be observed very efficiently on a 4-metre class
telescope because of their strong emission lines. We have added some
optical colour selection to our UV colour selection to further improve
the fraction of galaxies with high redshifts; the median redshift of
the measured galaxies is now $z_{med}=0.63$. The high sensitivity of
the new AAOmega spectrograph on the AAT and the short fibre
configuration time of the 2dF positioner mean that we can observe a
new field every 1.2 hours allowing us to make rapid progress on the
survey.

In this paper we have also presented our first public data release,
available from a dedicated web interface. The public data comprise
photometry (optical and UV), redshifts and spectra of all objects
observed up to the end of Semester 2008A. 

We cannot make the BAO clustering measurement at this stage of the
project. This is not only because we only have half the necessary
sample but also because the window function is very irregular due to
our highly variable coverage of the fields to date.  However, at the
half-way stage we already have the largest ever sample of high
redshift emission line galaxy spectra. Most importantly, we have
measured the small-scale clustering of these UV-luminous
galaxies and found it to be relatively strong which increases the
accuracy with which we will be able to measure the BAO scale. At lower
redshifts where we can measure the H$\beta$ line we have shown that
the WiggleZ galaxies are mostly extreme starburst-type galaxies as
expected. We have compared the WiggleZ galaxies to optically-selected
galaxies at the same redshift and confirmed that they are extremely
blue with high star formation rates. There is some evidence from {\em
  Hubble Space Telescope} imaging of a subset of the galaxies that the
starburst activity is being driven by interactions.

\section*{Acknowledgments}

This project would not be possible without the superb AAOmega/2dF
facility provided by the Anglo-Australian Observatory. We wish to
thank all the AAO staff for their support, especially the night
assistants, support astronomers and Russell Cannon (who greatly
assisted with the quality control of the 2dF system).

We also wish to thank: 
Alejandro Dubrovsky for writing software used to check the guide star
and blank sky positions;
Maksym Bernyk and David Barnes for help with the database
construction;
Peter Jensen and Max Spolaor for assistance with the redshift
measurements;
Michael Stanley for help with the selection of new GALEX positions; and
Michael Cooper for providing DEEP2 spectra for the
comparison in Sec.~\ref{sec-external}.

% Why these people are listed, for reference:
% 
% Chris Banks, UQ Hons - calibrating line fluxes (not used in this paper)
% Peter Jensen, UQ hons - redshifting
% Max Spolaor,  SUT PhD - redshifting
% Michael Stanley  UQ work experience  - GALEX tile centres
% Alejandro Dubrovsky --- UQ VO software person in LE grant

We wish to acknowledge financial support from The Australian Research
Council (grants DP0772084 and LX0881951 directly for the WiggleZ
project, and grant LE0668442 for programming support), Swinburne
University of Technology, The University of Queensland, the
Anglo-Australian Observatory, and The Gregg Thompson Dark Energy
Travel Fund.

GALEX (the Galaxy Evolution Explorer) is a NASA 
Small Explorer, launched in April 2003. We gratefully ac- 
knowledge NASA's support for construction, operation and 
science analysis for the GALEX mission, developed in co- 
operation with the Centre National d'Etudes Spatiales of 
France and the Korean Ministry of Science and Technology. 

Funding for the SDSS and SDSS-II has been provided by the Alfred
P. Sloan Foundation, the Participating Institutions, the National
Science Foundation, the U.S. Department of Energy, the National
Aeronautics and Space Administration, the Japanese Monbukagakusho, the
Max Planck Society, and the Higher Education Funding Council for
England. The SDSS Web Site is http://www.sdss.org/.

The RCS2 survey is based on observations obtained with 
MegaPrime/MegaCam, a joint project of CFHT and CEA/DAPNIA, at the
Canada-France-Hawaii Telescope (CFHT) which is operated by
the National Research Council (NRC) of Canada, the Institut National
des Sciences de l'Univers (CNRS) of France, and the University
of Hawaii.  The RCS2 survey is supported by grants to H.K.C.Y from
the Canada Research Chair program and the Discovery program of
the Natural Science and Engineering Research Council of Canada.

% This publication makes use of data products from the Two Micron All
% Sky Survey, which is a joint project of the University of
% Massachusetts and the Infrared Processing and Analysis
% Center/California Institute of Technology, funded by the National
% Aeronautics and Space Administration and the National Science
% Foundation.
% 
% Funding for the DEEP2 survey has been provided by NSF grants
% AST95-09298, AST-0071048, AST-0071198, AST-0507428, and AST-0507483 as
% well as NASA LTSA grant NNG04GC89G. Some of the data presented herein
% were obtained at the W. M. Keck Observatory, which is operated as a
% scientific partnership among the California Institute of Technology,
% the University of California and the National Aeronautics and Space
% Administration. The Observatory was made possible by the generous
% financial support of the W. M. Keck Foundation. 
% 
% Based on observations made with the NASA/ESA Hubble Space Telescope,
% and obtained from the Hubble Legacy Archive, which is a collaboration
% between the Space Telescope Science Institute (STScI/NASA), the Space
% Telescope European Coordinating Facility (ST-ECF/ESA) and the Canadian
% Astronomy Data Centre (CADC/NRC/CSA).

\bsp

\label{lastpage}

\end{document}